\RequirePackage{tikz}
\documentclass[sn-mathphys, iicol]{sn-jnl} 

\usepackage{amsmath}
\usepackage{amssymb}

\usepackage{subfig}
\usepackage{graphicx}
\usepackage{enumitem}

\usepackage{caption}
\usepackage[switch,pagewise]{lineno}

\DeclareMathOperator*{\argmax}{arg\,max}
\DeclareMathOperator*{\argmin}{arg\,min}

\newlength{\subcolumnwidth}

\newcommand{\nextsubcolumn}[1][]{%
    \cr\noalign{\hfill}
    \if\relax\detokenize{#1}\relax\else\hsize=#1\setlength{\subcolumnwidth}{\hsize}\fi
}

\jyear{2021}

\theoremstyle{thmstyleone}%
\theoremstyle{thmstyletwo}%

\theoremstyle{thmstylethree}%

\usetikzlibrary{intersections}
\usetikzlibrary{shapes.misc}
\tikzset{cross/.style={cross out, draw=black, fill=none, minimum size=2*(#1-\pgflinewidth), inner sep=0pt, outer sep=0pt}, cross/.default={2pt}}
\begin{document}
    
    \title[Geometry Parameter Estimation for Sparse X-ray Log Imaging]{Geometry Parameter Estimation for Sparse X-ray Log Imaging}
    
    \author*[1]{\fnm{Angelina} \sur{Senchukova}}\email{angelina.senchukova@lut.fi}
    
    \author[1]{\fnm{Jarkko} \sur{Suuronen}}\email{jarkko.suuronen@lut.fi}
    
    \author[2]{\fnm{Jere} \sur{Heikkinen}}\email{jere.heikkinen@finnos.fi}
    
    \author[1]{\fnm{Lassi} \sur{Roininen}}\email{lassi.roininen@lut.fi}
    
    \affil[1]{\orgdiv{School of Engineering Sciences}, \orgname{Lappeenranta--Lahti University of Technology LUT}, \orgaddress{\street{Yliopistonkatu 34}, \city{Lappeenranta}, \postcode{FI-53850}, \state{South Karelia}, \country{Finland}}}
    
    \affil[2]{\orgname{Finnos Oy}, \orgaddress{\street{Tukkikatu 5}, \city{Lappeenranta}, \postcode{FI-53900}, \state{South Karelia}, \country{Finland}}}

    \abstract{
         We consider geometry parameter estimation in industrial sawmill fan-beam X-ray tomography. 
         In such industrial settings, scanners do not always allow identification of the location of the source-detector pair, which creates the issue of unknown geometry. This work considers an approach for geometry estimation based on the calibration object. We parametrise the geometry using a set of 5 parameters. To estimate the geometry parameters, we calculate the maximum cross-correlation between a known-sized calibration object image and its filtered backprojection reconstruction and use differential evolution as an optimiser. The approach allows estimating geometry parameters from full-angle measurements as well as from sparse measurements. We show numerically that different sets of parameters can be used for artefact-free reconstruction. 
         We deploy Bayesian inversion with first-order isotropic Cauchy difference priors for reconstruction of synthetic and real sawmill data with a very low number of measurements. 
    }

    
    \maketitle

    \section{Introduction}
    
    A particular case of sparse X-ray computed tomography (CT) problems is found in the sawmill industry,  
    where X-ray scanners are employed 
    for non-invasive detection of log features and defects such as knots, rotten parts, and foreign objects. Furthermore, X-ray scanners allow timber tracking from the raw material to the resulting product (wood fingerprinting) \cite{Zol2019,Flod2008}. Non-destructive feature detection and timber tracking provide information for selection of optimal sawing strategies and efficient process control. 
     
    CT image reconstruction can be described mathematically as an inverse problem \cite{Tar2004,Kaip2005,Stu2010}.  
    The objective of two-dimensional X-ray CT is to reconstruct the cross-sectional image of a scanned object from a collection of one-dimensional line projections. The projections are obtained by measuring the intensity loss of the beam of X-rays that penetrates the object from different view angles. The projections are collected to a sinogram. 

    In the sawmill industry, highly limited (sparse) tomographic data is used for log imaging. The sparse tomography is known to be a severely ill-posed problem. Sparse data lead to severe artefacts when the image is reconstructed by conventional methods like filtered backprojection~\cite{Nat1986}. 
    
    To solve ill-posed inverse problems, a Bayesian approach is often applied \cite{Stu2010}. The Bayesian framework naturally allows for incorporating a prior distribution which encodes the possible internal structure of the object. First-order isotropic Cauchy difference priors are shown to promote sharp edges, which enables detection of sharp features and material interfaces with a very low number of measurements in the sawmill applications~\cite{Suur2020}.
    
    The CT scanner geometry describes the location and orientation of the source-detector pair for each projection of the CT scan with respect to the reference system of the scanned object. 
    Geometric inaccuracies are known to cause errors and severe artefacts in reconstructions \cite{LiGull1994,Huili1998}.
    In practical industrial settings in sawmills, the structure of an X-ray scanning device does not always allow for the direct measurement of the important geometry parameters such as the distance between the source and detector or the detector tilt. 
    Moreover, geometric errors can arise due to factors such as vibration during motion of either the scanner or scanned object \cite{Jereb2012}. There is, therefore, a need to address the issue of unknown geometry parameters.

    In this work, we estimate the fan-beam geometry defined by 5 parameters: the first scanning angle, the distance between the centre of rotation and the detector, the source shift, the detector shift, and the detector tilt. In the particular application to sawmill log imaging discussed in the paper, a single source-detector pair is fixed while the scanned object is rotated to obtain its projections at different angles. 
    
    \begin{figure}[]
        \centering
        \includegraphics[width=0.9\linewidth]{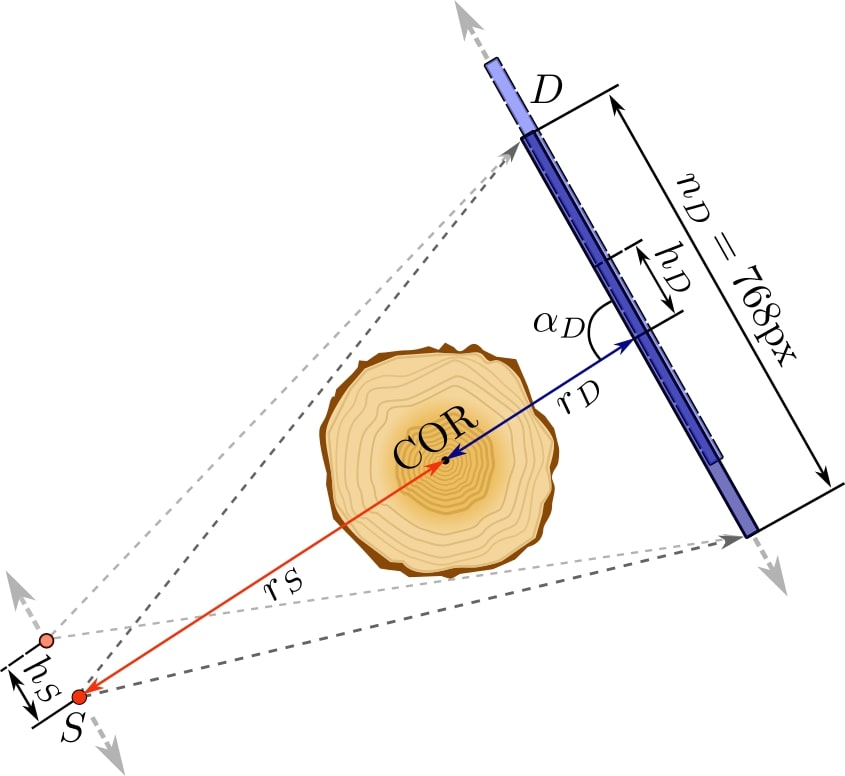}
        \caption{A scheme of the X-ray measurement geometry. The point source $S$ and the flat-line detector $D$ are placed around the log. The log is rotated around the centre of rotation COR to obtain different projections. The distance between the source and COR is $r_S$ and the distance between COR and the detector is $r_D$. In the geometry, there are possible shifts of source $h_S$ and detector $h_D$ in tangent directions, and the detector tilt $\alpha_D$ }\label{fig:x-ray_setup}
    \end{figure}

    \subsection{Related Work}
    
    Existing studies on geometry calibration can be roughly divided into two groups: 1) methods using calibration phantoms, and 2) self-calibration (marker-free) methods. Calibration phantom-based methods, which are the more common approach, use one or more calibration objects (calibration phantoms) incorporating a~known arrangement of radiopaque markers such as steel ball bearings. An extensive review of such approaches is given in \cite{Ferr2015}. An overview of marker-free methods for estimating the crucial geometry parameter, the axis of rotation, is given in \cite{Zem2023}.
    
    A pioneering study \cite{Gull1987} on imaging system geometry estimation utilises single photon emission computed tomography. 
    In the study, the Levenberg-Marquardt optimisation algorithm is used to calibrate a fan-beam geometry from projection measurements of reference objects consisting of point markers.     
    The same principles are also appropriate for calibrating X-ray CT systems.
    
    A calibration method for  cone-beam geometry that employs a calibration phantom to estimate the geometry parameters is proposed in \cite{Cho2005}. The method uses a transparent cylindrical acrylic tube with 24 steel ball bearings as a calibration phantom, but the approach can be adapted also to other phantom configurations. 
    
    A self-calibration method is introduced in \cite{Ouadah2016}, where calibration is done in the cone-beam setting without calibration objects. 
    The method performs calibration by registering two-dimensional projection data to a previously acquired three-dimensional image of the scanned object.

    Joint estimation of uncertain view angles determining the geometry of the forward CT model and the attenuation coefficient function of a scanned object is discussed in \cite{Uribe2021,Hansen2021}. In~\cite{Ped2023}, the geometry estimation, in particular, the center-of-rotation offset, is performed jointly with reconstruction based on a Bayesian approach that also provides information on the uncertainty of geometry parameters.

    In \cite{Gen2021}, data-driven estimation of unknown fan-beam geometry is done by employing a neural network that learns the unknown forward operator from training data consisting of sinogram-image pairs. 
    Another learned approach to tackle  parametric uncertainties in the measurement operator is introduced in \cite{Xie2020}. The approach allows for joint automatic geometry parameter calibration along with reconstruction of the unknown CT image.

    \subsection{Our Contributions}
    
    We propose a new approach for identifying unknown geometry parameters in fan-beam X-ray log tomography by using a calibration object of known size that is scanned jointly with a log. 

    Compared to other existing works on geometry calibration, the difference in this approach is that: 1) it employs an easy-to-produce phantom for calibration of X-ray CT geometry with a fixed source-detector pair and rotating scanned object; 2) the approach itself is based on calculating the maximum cross-correlation between the ground truth image of the calibration object and its filtered backprojection (FBP) reconstruction; and 3)~the approach allows estimating geometry parameters even for a limited number of projection angles. 

    Additionally, we demonstrate a number of possible artefacts that may appear in the reconstructions in the eventuality of misspecified geometry parameters when using a fan-beam configuration with a flat line detector. 
    In extreme cases, misspecifications can lead to full unidentifiability, and in less severe cases, the reconstructions can contain duplicated features, halos, shifts, and blurring.
    We aim to demonstrate these features and present a~way to overcome the issue with suitable parameter choices.
    
    Finally, we consider the sparse-angle tomography and employ maximum a posteriori estimates with a recently developed class of priors, first-order isotropic Cauchy difference priors \cite{Suur2022}. We demonstrate the amenability, for our specific use case, of this method over conventional methods (filtered backprojection, Tikhonov regularisation) when the number of projection data is extremely limited.

    \subsection{Outline of the Paper}
    
    The rest of this paper is organised as follows: 
    In Section \ref{sec:tomo}, we review the tomography reconstruction model and methods.
    The new geometry parameter estimation algorithm is presented in Section \ref{sec:geometry}.
    Synthetic and real data CT examples are demonstrated in Section \ref{sec:results}. 
    Section \ref{sec:discussion} concludes the study.

    \section{Tomography reconstruction}
    \label{sec:tomo}

    \subsection{Fan-beam geometry setting}
    
    The schematic plot of the X-ray scanner geometry used in the numerical experiments is shown in Figure~\ref{fig:x-ray_setup}. An X-ray source, placed around the log, emits fan-shaped X-ray beams that are recorded by the corresponding flat line detector. 
    
    The geometry of the fan-beam model is determined by seven parameters in total.  There are two known parameters:
    \begin{itemize}
        \item $r_{S} = 859.46$ (mm) --- distance between the X-ray source and centre of rotation (COR), and
        \item $n_{D} =768$ (px) --- number of detector elements, the pixel size is $2$mm,
    \end{itemize}
    and the other five parameters are treated as unknown:
    \begin{itemize}
        \item $\alpha_{0}$ (rad) --- the first scanning angle; the projection angles are defined w.r.t. the first scanning angle, and are assumed to be uniformly distributed between the first and last angles,
        \item $r_{D}$ (mm) --- distance between the COR and detector,
        \item $h_{S}$ (mm)--- the source shift (in the tangent direction),
        \item $h_{D}$ (mm)--- the detector shift (in the tangent direction), and
        \item $\alpha_{D}$ --- defines the detector tilt. In the default configuration, the initial source-to-detector vector is $[0, 1]$, and the initial detector axis is perpendicular to it (i.e. $[1, \alpha_{D} = 0]$). However, if the detector is tilted then component~$\alpha_{D}$ should be adjusted accordingly.
    \end{itemize}

    We further denote those five unknown parameters as a 5-dimensional vector $$\boldsymbol\theta = [\alpha_0, r_{D}, h_{S}, h_{D}, \alpha_D]^T. $$

    \subsection{Discrete CT imaging model}

    Using the Beer-Lambert law \cite{Nat1986}, the intensity loss of X-ray beams passing through a scanned object can be represented as
    

    \begin{equation}\label{eq:beer-lambert}
        \frac{I_{\varphi, \tau}}{I_0} = \exp\left(- \int_{l_{\varphi, \tau}} x(s) ds\right),
    \end{equation}
    where $x$ is the attenuation coefficient describing the internal structure of the object, $I_0$ and $I_{\varphi, \tau}$ are the initial and final intensities, respectively, and $l_{\varphi, \tau}$ denotes the line corresponding to the acquisition angle $\varphi \in [0, 2\pi)$ and detector location $\tau \in \mathbb{R}$. 

    Taking logarithms in Equation (\ref{eq:beer-lambert}) we obtain the following 

    \begin{equation}\label{eq:beer-lambert_ln}
        -\ln \left(\frac{I_{\varphi, \tau}}{I_0} \right)= \int_{l_{\varphi, \tau}} x(s) ds,
    \end{equation}
    where the left-hand side $y := -\ln(I_{\varphi, \tau}/I_0)$ is measured data known as projection. The collection of projections is commonly referred to as a sinogram. The right-hand side of (\ref{eq:beer-lambert_ln}) is known as the Radon transform 
    
    \begin{equation*}
        \mathcal{R}(x) := \int_{l_{\varphi, \tau}} x(s) ds,
    \end{equation*}
    that maps the unknown attenuation coefficient function $x$ into the set of its line integrals \cite{Rad1986}. In other words, the known measured data is represented by the Radon transform, which corresponds to the forward CT model.
    



    In practice, only a finite number of X-ray projections is available and the sinogram measurements are recorded by the pixel detectors. If the number of acquisition angles is $K$, the number of detector pixels is $M$, and a scanned object is discretised into a $N \times N$ grid, we obtain the discrete model for the X-ray CT imaging
    
    \begin{equation}\label{eq:forward_model}
    \mathbf{y} = \mathbf{A}_{\boldsymbol{\theta}}\mathbf{x} + \mathbf{e},
    \end{equation}
    where $\mathbf{x} \in \mathbb{R}^{N^2\times 1}$ is the unknown vector of attenuation coefficients, $\mathbf{A}_{\boldsymbol{\theta}} \in \mathbb{R}^{KM\times N^2}$ is a so-called system matrix (forward matrix) describing intersection lengths for each of $KM$ lines. The system matrix is assumed constant throughout the reconstruction and depends on the scanning geometry parameters $\boldsymbol{\theta}$ that are not known exactly. Vector $\mathbf{y}\in \mathbb{R}^{KM\times 1}$ specifies the measurement data (projections) taken during the scanning process, and $\mathbf{e}$ denotes measurement noise. 
    
    %
    %

    The corresponding inverse problem aims to reconstruct $\mathbf{x}$ from the available measured sinogram data $\mathbf{y}$. This can be done either in Bayesian estimation or a classical deterministic framework~\cite{Kaip2005}. In the paper, we use and compare three different reconstruction methods (filtered backprojection, Tikhonov regularisation and MAP estimate) discussed in Section~\ref{sec:reco_methods}.

    \subsection{Reconstruction methods}
    \label{sec:reco_methods}


    The traditional method to reconstruct the cross-sectional image of an object from the measurement data  is \textit{filtered backprojection (FBP)}. 
    If $\mathbf{y}$ denotes the measurement data and $\boldsymbol\theta$ is the geometry parameter vector, then the discretised form of the FBP formula is defined as  
    
    \begin{equation}\label{eq:fbp_discr}
    \mathbf{x}_{\mathrm{FBP}} = \frac{1}{2} \mathbf{B}_{\boldsymbol \theta} (\boldsymbol\phi \bar{\ast} \mathbf{y}) =: \mathrm{FBP}(\boldsymbol\theta; \mathbf{y}), 
    \end{equation}
    where $\boldsymbol{\phi}$ is a convolutional filter in the projection domain, $\mathbf{B}_{\boldsymbol \theta}$ is discrete backprojection, and $\bar{\ast}$ denotes the discrete convolution. 
    
    Several frequency domain filters are commonly used in FBP, including the most basic Ram-Lak filter \cite{Kak2001} and its windowed modifications such as Shepp-Logan, Cosine, Hamming and Hann filters. The filter function significantly affects the robustness of the reconstruction and should be carefully selected based on the particular reconstruction task.
 
    In sparse CT where the number of projections is extremely limited, the FBP method leads to severe artefacts in reconstructions due to ill-posedness of the problem \cite{Nat1986}. 
    Different regularisation techniques have been used to solve ill-posed problems. One of the most common regularisation techniques, \textit{Tikhonov regularisation}, is defined as a solution of the following minimisation problem 
    
    \begin{equation}\label{eq:tikh_reg}
    \mathbf{x}_\mathrm{Tikh} = \argmin_{\mathbf{x}} \left\{ \Vert \mathbf{A}_{\boldsymbol{\theta}}\mathbf{x} - \mathbf{y}\Vert^2 + \alpha \Vert\mathbf{x}\Vert^2 \right\}.
    \end{equation}
    The regularisation parameter  $\alpha > 0$ is chosen depending on the noise strength to find a trade-off between the solution stability and its ability to explain the measurements.

    In the Bayesian formulation of inverse problem~(\ref{eq:forward_model}), the observed data $\mathbf{y}$, the unknown parameter $\mathbf{x}$ and noise $\mathbf{e}$ are treated as random variables. The noise $\mathbf{e}\sim N(\mathbf{0},\mathbf{\Sigma})$ is  assumed to be a zero-mean Gaussian multivariate with covariance matrix $\mathbf{\Sigma}$.
    
    According to Bayes' theorem, the posterior distribution of unknown $\mathbf{x}$ can be formulated as 
    
    $$\pi_\text{post}(\mathbf{x}) = \pi(\mathbf{x} \vert \mathbf{y}) = \frac{\pi_\text{pr}(\mathbf{x}) \pi(\mathbf{y} \vert \mathbf{x})}{\pi(\mathbf{y})}.$$
    The prior distribution $\pi_\text{pr}(\mathbf{x})$ incorporates a~priori information about the unknown $\mathbf{x}$ before the measurements were obtained, and $\pi(\mathbf{y} \vert \mathbf{x})$ is the data-likelihood that contains information about the forward operator $\mathbf{A}_{\boldsymbol{\theta}}$ and assumptions about the statistical distribution of noise $\mathbf{e}$ \cite{Kaip2005,Stu2010}. 


    The features on the boundaries between different regions are of particular interest in reconstructions. In sawmill applications, sharp-interface reconstructions are needed for measuring the size and shape of tree knots. Various edge-preserving regularisation techniques have been proposed to avoid smoothing of edges in CT \cite{Yu2002,Xu2019,Hama2013}. 
    
    As an edge-preserving method, we employ \textit{the maximum a posteriori (MAP) estimate} of the posterior distribution 

    \begin{equation}\label{eq:map_estim}
    \mathbf{x}_\text{MAP} = \argmax_{\mathbf{x} \in \mathbb{X}} \pi_\text{post}(\mathbf{x}), 
    \end{equation}
    with the first-order isotropic Cauchy difference prior \cite{Suur2020,Suur2022, Markkanen19} and  the following data likelihood term 
    
    \begin{equation*}
    \pi(\mathbf{y}\vert \mathbf{x}) \propto \exp\left(-\frac{1}{2} \Vert \mathbf{A}_{\boldsymbol{\theta}}\mathbf{x} - \mathbf{y}\Vert^2\right). 
    \end{equation*}
    
    Unlike  the standard Tikhonov regularisation~\eqref{eq:tikh_reg}, which assumes no special structure for $\mathbf{x}$ except that it should be close to zero, the first-order Cauchy difference prior incorporates prior information on how the first-order derivatives of $\mathbf{x}$ will be distributed. Employing Cauchy distribution for the distribution of the differences effectively allows the existence of discontinuities in $\mathbf{x}$ by favouring features that are piecewise close to constant. 
    
    If $\mathbf{x}$ is assumed to be discretised in a two-dimensional uniform grid indexed by $(i,j)$ with size $N \times N$, then the isotropic Cauchy difference prior distribution can be expressed as follows \cite{Suur2022}
    
    \begin{equation}
    \label{eq:ci1}
    \begin{split}
    \pi_{\textrm{pr}}(\mathbf{x}) & \propto 
    \prod_{i=1}^{N-1} \prod_{j=1}^{N-1} \big( \beta^2 + (x_{i+1,j}  - x_{i,j})^2  + \\
    & (x_{i,j+1} - x_{i,j})^2 \big)^{-3/2},
    \end{split}
    \end{equation} 
    where $\beta^2$ is a parameter that controls the strength of the prior in terms of how close to zero the first-order differences should be. 
    

     \begin{figure}[]
        \centering
        \includegraphics[width=1\linewidth]{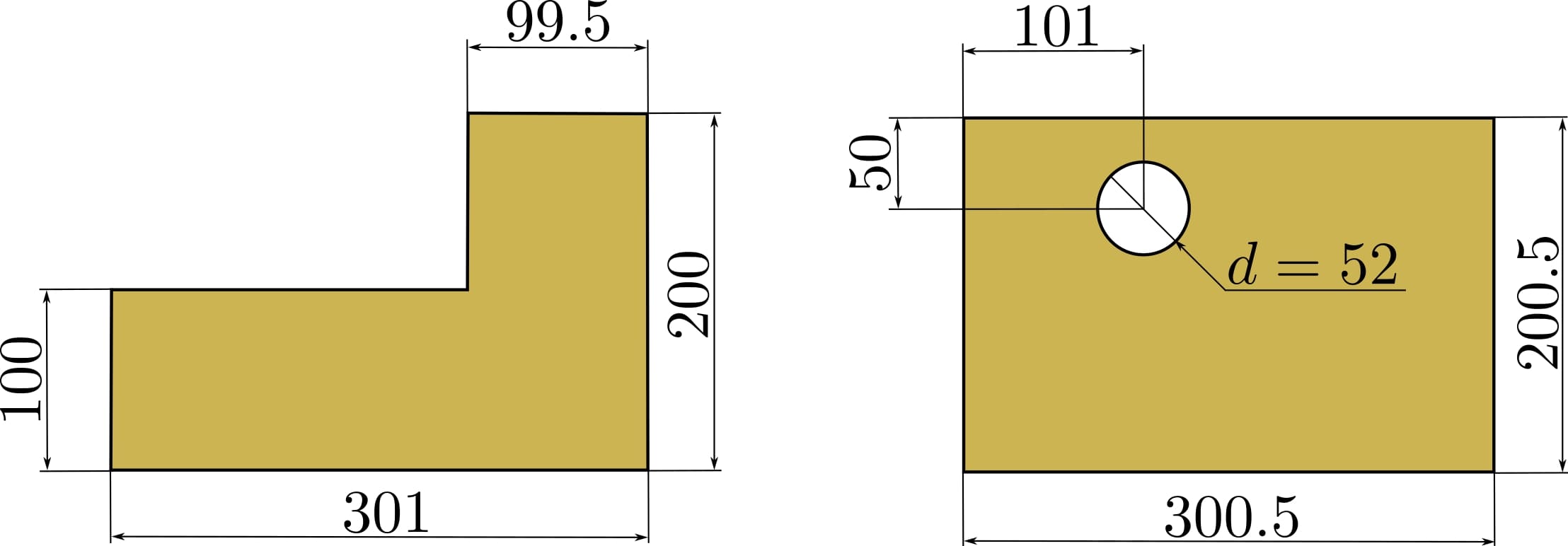}
        \caption{Images of the L-shaped calibration phantom (left) and the calibration phantom with a hole (right). All dimensions are given in mm} \label{fig:calibration_phantoms}
    \end{figure}

    \section{Geometry parameter estimation}
    \label{sec:geometry}
   
    For geometry calibration, we use two different calibration phantoms of known size and shape: an L-shaped calibration phantom and a calibration phantom with a hole (see Figure~\ref{fig:calibration_phantoms}). 
    
    \subsection{Geometry estimation model}
    The parameters of the partially unknown geometry of the measurement equipment  are estimated using an objective function that evaluates the correlation of a reference image to an FBP reconstruction with given parameters. More precisely, we compare both the reference image and its mirrored image to the reconstruction, because the ambiguity of the direction of the detector plates and the rotation direction of the object inside the scanning device may render the reconstructions mirrored as well. The ambiguity of the handedness of the reconstructions can  be taken into account in the objective function by calculating the maximum of correlations of both handednesses. 
    
    We seek to find a vector of unknown geometry parameters $\boldsymbol{\theta}$ based on the calibration phantom image $\mathbf{X}_\mathrm{ref}$ of $N\times N$ pixels and its mirrored image $\widehat{\mathbf{X}}_\mathrm{ref}$. To separate the reference image from its flattened vector form $\mathbf{x}_\mathrm{ref}$, we denote the image ($N\times N$ pixels) with the capitalised $\mathbf{X}_\mathrm{ref}$. 
    
    Let $ \mathbf{f}_{\boldsymbol{\theta}} = \mathrm{FBP}(\boldsymbol{\theta};\mathbf{y})$ be the corresponding FBP reconstruction  that depends on the input X-ray (sinogram) data $\mathbf{y}$  and geometry parameters $\boldsymbol{\theta}$. 
    Then the objective function $J(\boldsymbol{\theta})$ is defined as the negative maximum of the correlation between  $\mathbf{X}_\mathrm{ref}$ and $\mathbf{f}_{\boldsymbol{\theta}}$  
    and the correlation between  $\widehat{\mathbf{X}}_\mathrm{ref}$ and $\mathbf{f}_{\boldsymbol{\theta}}$
    
    
    \begin{equation}\label{eq:obj_func}
    J(\boldsymbol{\theta}) = -\max\left\{ \mathrm{Tr}\left( {\mathbf{X}}_\mathrm{ref}^T \mathbf{f}_{\boldsymbol{\theta}}\right), \mathrm{Tr}\left(\widehat{\mathbf{X}}_\mathrm{ref}^T \mathbf{f}_{\boldsymbol{\theta}}\right)  \right\},
    \end{equation}
    where $\mathrm{Tr}$ means the trace of a matrix, and the superscript $T$ means transpose. In other words,  the objective function evaluates the Frobenius inner product of the FBP reconstruction with given geometry parameters and two possible calibration object candidates. It returns the negated maximum of the values. In this study, we do not employ the normalised correlation in the method, because the unnormalised correlation proved effective and robust. For the second, usage of the normalised correlation requires addressing the exceptional case when the standard deviation of the reconstruction is zero. This might be the case when the differential evolution proposes a set of geometry parameters that results in an empty reconstruction in the domain of the reference object, although such a case is very unlikely.    
    
    We also experimented with an objective function based on sinogram  data directly rather than FBP,  namely 
    
    \begin{equation}\label{eq:alt_obj_func}
    \tilde{J}(\boldsymbol{\theta}) = -\max\left\{ \mathrm{}\left( \mathbf{A}_{\boldsymbol{\theta}} {\mathbf{x}}_\mathrm{ref}\right) ^T  \mathbf{y}, \mathrm{}\left(\mathbf{A}_{\boldsymbol{\theta}} \widehat{\mathbf{x}}_\mathrm{ref}\right) ^T  \mathbf{y}   \right\},
    \end{equation}
    where $\mathbf{A}_{\boldsymbol{\theta}}$ is the forward CT operator matrix that depends on the geometry parameters $\boldsymbol\theta$. However, the objective function (\ref{eq:alt_obj_func}) demonstrated worse performance in comparison with the objective function (\ref{eq:obj_func}). More specifically, FBP reconstructions using the set of found geometry parameters contained severe artifacts and were often misaligned with the reference object. We argue this is because the optimisation schemes cannot propose and find good sets of geometry parameters to align the reference and measured sinograms when the correlation is performed in the projection space, where the differences between the perturbed and unperturbed geometry parameters appear relatively subtle.  Instead, the effect of perturbation of the geometry parameters is typically more substantial in the reconstruction space, which helps the optimiser to find the potential global optimum of the cross-correlation, which, in turn, ensures the proper alignment of the geometry.   
    
    From a numerical viewpoint, simultaneous evaluation of two calibration object scenarios in the objective function  (\ref{eq:obj_func}) may pose additional challenges to the optimisation algorithms as the likelihood of getting trapped to a local minimum might increase. Evaluating the two calibration object cases separately is plausible. However, even an objective function that evaluates the correlation with one calibration object at times is non-differentiable,  so we use the objective function presented in Equation (\ref{eq:obj_func}) for the unified optimisation workflow.

    \subsection{Differential Evolution}

    The aim of geometry parameter estimation task is to find parameters $\boldsymbol{\theta}^*$  that minimise the objective function  $J(\boldsymbol{\theta})$
    
    \begin{equation}\label{eq:opt_prob}
    \boldsymbol{\theta}^* = \argmin_{\boldsymbol{\theta}} J(\boldsymbol{\theta}).
    \end{equation}
    To find a solution to the optimisation problem~\eqref{eq:opt_prob},  we use the differential evolution (DE) method \cite{StornPrice2020}. The method belongs to evolutionary algorithms that apply natural mechanisms of evolution theory such as mutation and selection to evolve a solution to an optimisation problem. 
    
    Let $S \subset \mathbb{R}^D$ be the search space of the optimisation problem to be solved. 
    Then the DE algorithm keeps a population of $N_\mathrm{pop}$ individuals ($D$-dimensional vectors) 
    
    \begin{equation*}
    \mathbf{x}_i = [x_{i, 1}, x_{i, 2}, \ldots, x_{i, D}]^{T}, \quad i=1, \ldots, N_\mathrm{pop},
    \end{equation*}
    each of these individuals stands for a possible solution of the problem.  At every pass through the population, the algorithm mutates each candidate solution by other candidate solutions to create a mutant candidate. 
    
    Different mutation strategies are used in DE \cite{Red2007,Leon2014}. In this work, we adhere to the DE/best/1/bin strategy and use its implementation in the SciPy Python library \cite{SciPy2020} (the pseudo-code is listed in Algorithm~\ref{alg:diff_evol}). In this strategy, the best member of the population, $\mathbf{x}_{\mathrm{best}}$,  is mutated (the  \textbf{Mutation}  stage) by the difference of two other randomly selected members $\mathbf{x}_{k_1}$ and $\mathbf{x}_{k_2}$, ($k_1 \neq k_2 \neq \mathrm{best}$)  to form a mutant vector $\mathbf{m}_i$ as follows
    
    \begin{equation*}
    \mathbf{m}_i = \mathbf{x}_{\mathrm{best}} + \mu (\mathbf{x}_{k_1} - \mathbf{x}_{k_2}),
    \end{equation*}
    where $\mu \in [0, 2]$ is the mutation constant that controls the amplification of the differential variation $(\mathbf{x}_{k_1} - \mathbf{x}_{k_2})$. 

    \begin{figure}[]
        \centering
        \includegraphics[width=0.5\linewidth]{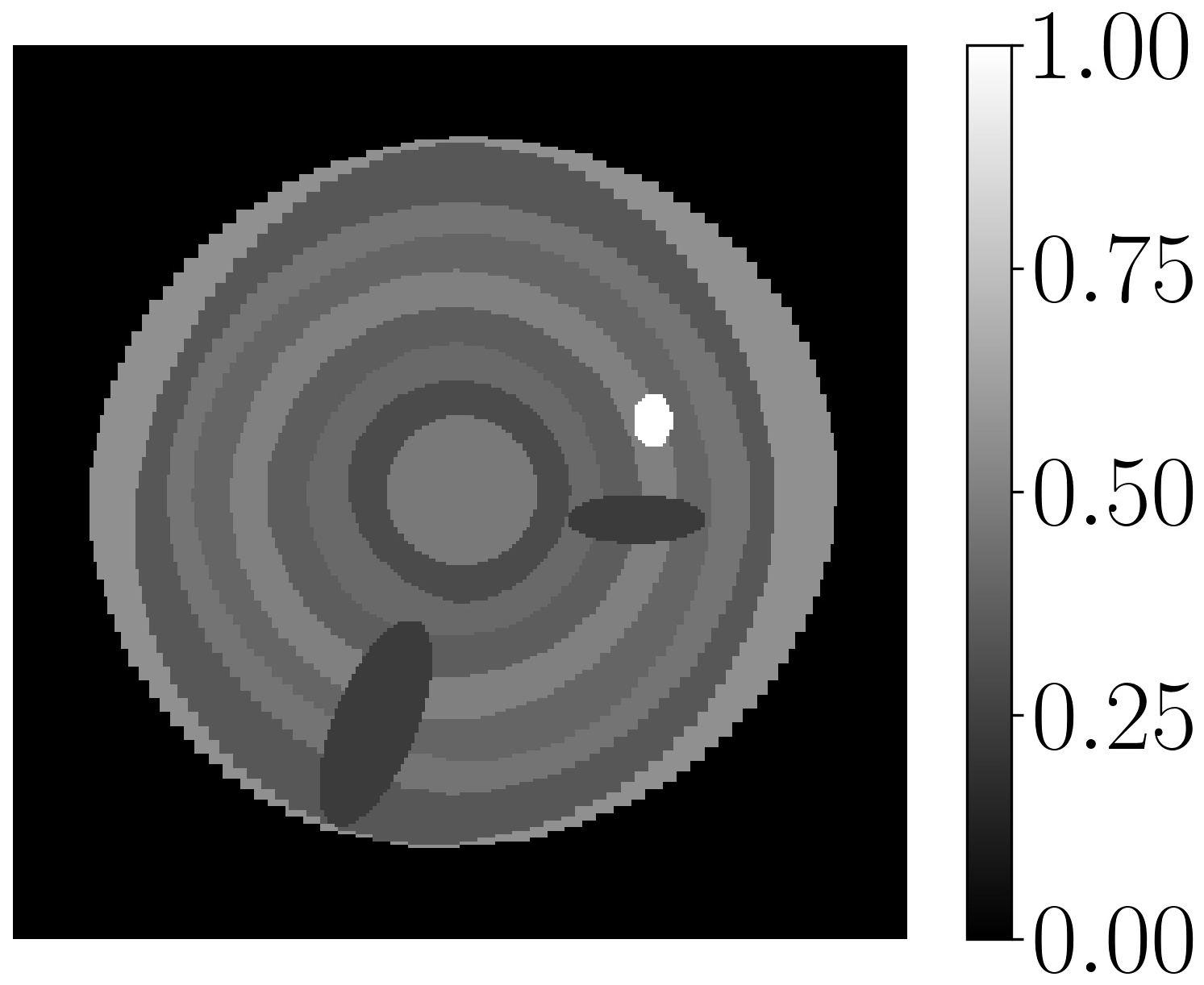}
        \caption{A digital log phantom representing the cross-sectional image of a tree log (image resolution $256 \times 256$ pixels) }\label{fig:log_phantom}
    \end{figure}

    Next, at the \textbf{Crossover} stage, the trial vector is formed as follows
    
    \begin{equation*}
    u_{i, j} = \begin{cases} m_{i, j}, & \mathrm{if} \; r_{i, j} < P_\mathrm{cross},\\ 
    x_{i, j}, & \mathrm{otherwise},  \end{cases}
    \end{equation*}
    for $j=1,  \ldots,  D-1$, where $r_{i,j} \sim U(0,1)$ is a random number drawn from the uniform distribution on the interval $[0,1]$, and $P_\mathrm{cross} \in [0, 1]$ is a crossover (recombination) probability \cite{StornPrice2020}. In the numerical experiments, we use $P_\mathrm{cross} = 0.7$. The final entity $u_{i, D}$ is always calculated as 
    $$u_{i, D} = m_{i, D}.$$ 
    
    Eventually, at the \textbf{Selection} stage, the fitness of the trial candidate is assessed: if the trial candidate is better than the original candidate then it takes its place.

    \captionsetup[subfigure]{labelformat=empty, position=top}
    \begin{figure*}[]
        \centering
        \includegraphics[width=0.9\linewidth]{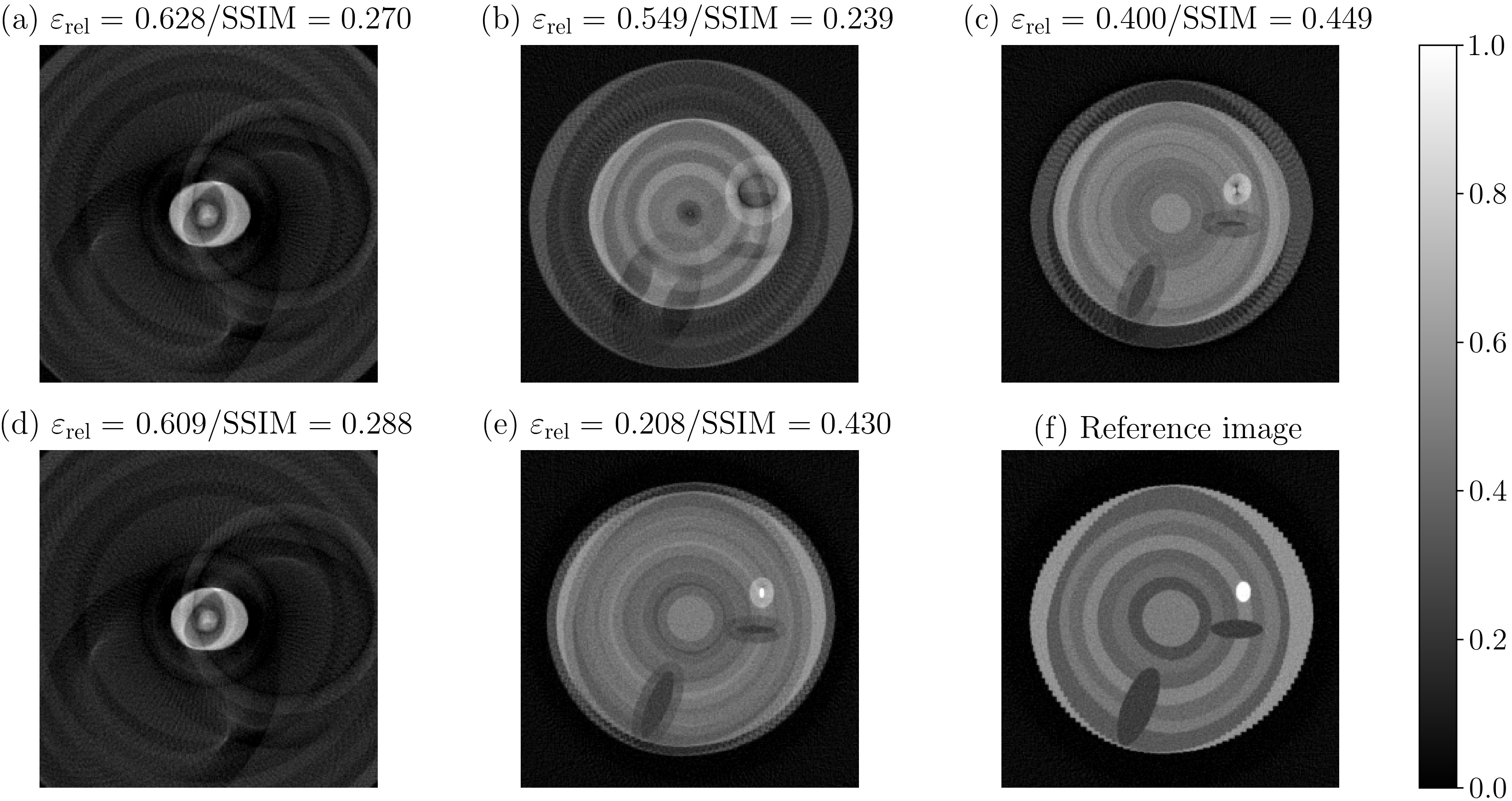}
        \caption{Example artefacts in reconstructions when geometry parameters are incorrectly specified: (a)~the source and detector shifts, detector tilt are misspecified; (b)~the detector shift and tilt are misspecified; (c)~the detector tilt is misspecified;  (d)~the detector and source shifts are misspecified; (e)~the detector radius is slightly perturbed; (f)~all the parameters are specified correctly. FBP reconstruction~(f) with true parameters is used as a reference for computing  $\varepsilon_{\mathrm{rel}}$ and SSIM} \label{fig:ph_bad_recos}
    \end{figure*}
    
    \begin{algorithm}[]  
        \caption{Differential Evolution (DE)}\label{alg:diff_evol}
        \begin{algorithmic}[1]
            \State \textbf{Input parameters:} $D$ -- problem dimensionality; $N_\mathrm{pop}$ -- population size; $\mu$ -- mutation constant; $P_\mathrm{cross}$ -- crossover probability; $f$ -- objective function under consideration;
            \State \textbf{Initialisation: } Create population with $N_\mathrm{pop}$ individuals and initialise them with random positions in the search space;
            \While{the termination condition is not satisfied}
            \For{$i = 1..N_\mathrm{pop}$}
            \State \textbf{Generation}: Take the best individual \State $\mathbf{x}_{\mathrm{best}}$ in the current population;
            \State Choose 2 random integers 
            \State $k_1$, $k_2\in (1, N_\mathrm{pop})$  such that 
            \State $k_1 \neq k_2 \neq \mathrm{best}$;
            \State \textbf{Mutation:} Form a mutant vector 
            \State $\mathbf{m}_i \gets \mathbf{x}_{\mathrm{best}} + \mu (\mathbf{x}_{k_1} - \mathbf{x}_{k_2})$;
            \State \textbf{Crossover:}
            \For{$j = 1..D-1$}
            \State Generate $r_{i, j} \sim U(0,1)$;
            \If{$r_{i, j} < P_\mathrm{cross}$}
            \State $u_{i, j} \gets m_{i, j}$;
            \Else
            \State $u_{i, j} \gets x_{\mathrm{best}, j}$;
            \EndIf
            \EndFor
            \State $u_{i, D} \gets m_{i, D}$;
            \State \textbf{Selection:}
            \If{$f(\mathbf{u}_i) \leq f(\mathbf{x}_{\mathrm{best}})$}
            \State $\mathbf{x}_{\mathrm{best}} \gets \mathbf{u}_i$;
            \EndIf
            \EndFor
            \EndWhile
        \end{algorithmic}
    \end{algorithm}

    The choice of the DE method for our problem of geometry parameter optimisation  stems from the fact that DE is able to cope with  nonlinear and non-differentiable objective functions and requires only a few control variables. It is also a global optimisation algorithm, which is  important for our geometry parameter search. There are several local optima  in the parameter space when evaluating the correlation between a reference object and the parametrised FBP reconstructions. Consequently, local optimisation algorithms that do not require the objective function to be differentiable, such as the Nelder-Mead method \cite{Nelder1965}, are not suitable for this task.  It must be recalled that the objective function likely does not even have a unique optimum: there might be multiple parameter vectors that give almost the same cross-correlation with the reference object when used in the FBP reconstruction function. However, the uniqueness of the geometry parameters of the measurement equipment is not interesting as such, since we are only interested in the quality of reconstructions.

    In addition, we claim that the nature of the objective function \eqref{eq:obj_func}  favors DE over simulated annealing (SA), which is  another common global optimisation algorithm \cite{SA1989,Horst2013}. Differential evolution utilises a set of particles  that mimic the natural selection, evolution and mutation of the best candidates for the global maximum of the  objective function. When some of the geometry parameter candidates have been mutated closer to a possible optimum, the evolution steps help to find the precise maximum in a more robust manner than SA. 
    

    \section{Results and discussion} 
    \label{sec:results}

    In the numerical experiments, we used the Operator Discretization Library (ODL) \cite{JonasAdler2017}.  ODL is a Python library developed at KTH Royal Institute of Technology, Stockholm. The library focuses on inverse problems including CT reconstruction.

     \begin{figure*}[]
        \centering
        \includegraphics[width=0.9\linewidth]{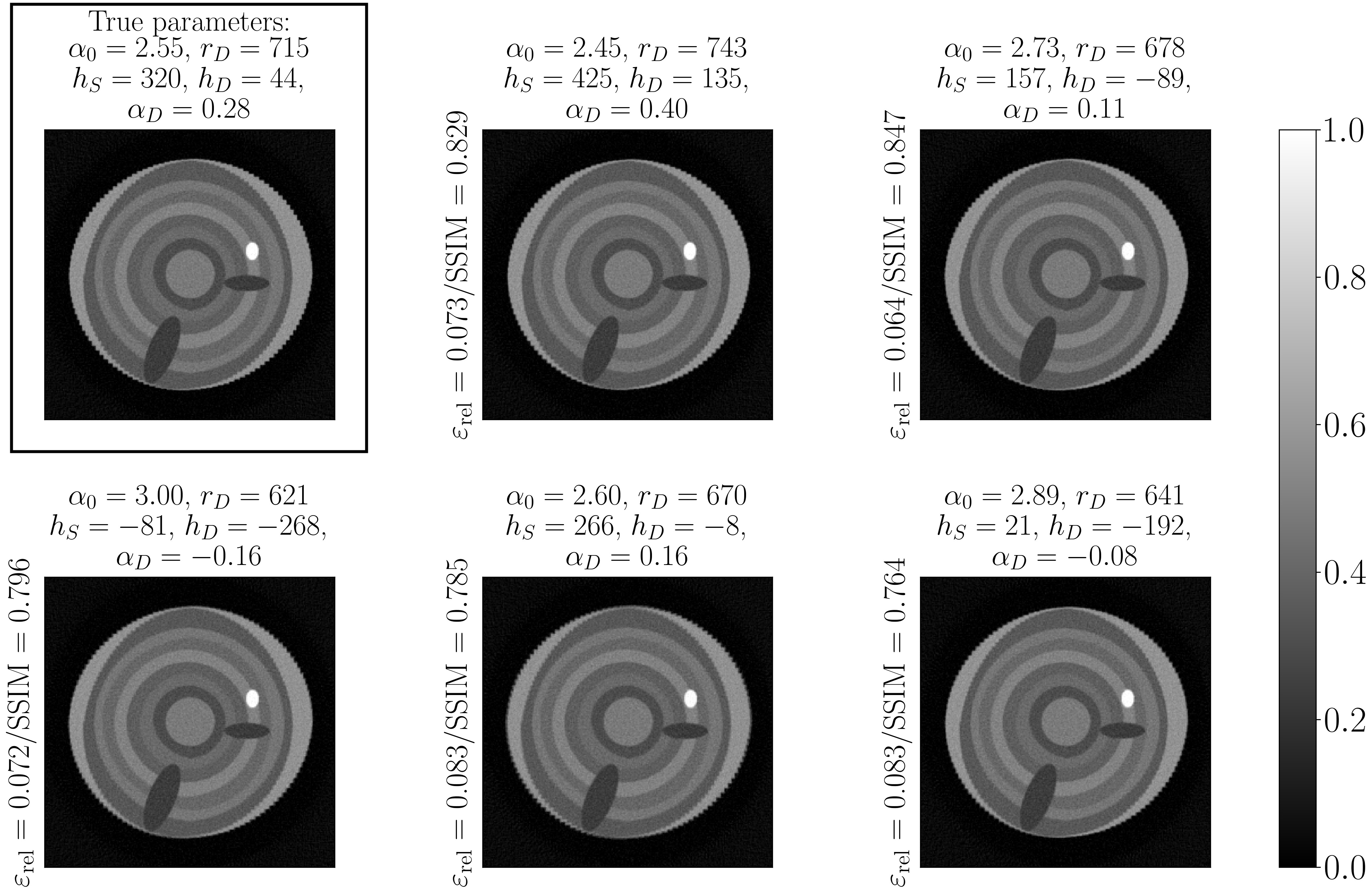}
        \caption{Full-angle FBP-reconstructions of log phantom (image resolution $256 \times 256$ pixels): true geometry parameters (framed subplot) and different parameter vectors found by the calibration method with the L-shaped phantom. FBP reconstruction with true parameters is used as a reference for computing  $\varepsilon_{\mathrm{rel}}$ and SSIM}\label{fig:phantom_diff_parametrisations}
    \end{figure*}  

    \begin{table*}[]
    \centering
    \begin{tabular}{c|rrrrr|c|c}
    \hline
                     & \multicolumn{5}{c|}{Geometry parameters}                                                                                                             &  &  \\
                     & \multicolumn{1}{c}{$\alpha_0$} & \multicolumn{1}{c}{$r_D$} & \multicolumn{1}{c}{$h_S$} & \multicolumn{1}{c}{$h_D$} & \multicolumn{1}{c|}{$\alpha_D$} & $\varepsilon_{\mathrm{rel}}$                             &    SSIM  \\
& \multicolumn{1}{c}{(rad)}      & \multicolumn{1}{c}{(mm)}  & \multicolumn{1}{c}{(mm)}  & \multicolumn{1}{c}{(mm)}  & \multicolumn{1}{c|}{}           &                              &      \\
\hline
True parameters                & $2.55$                         & $715$                     & $320$                     & $44$                      & $0.28$                         &                              &      \\
\hline
\multicolumn{1}{c|}{} & $2.45$                         & $743$                     & $425$                     & $135$                     & $0.40$                         & $0.073$                      &  $0.829$    \\
                     & $2.73$                         & $678$                     & $157$                     & $-89$                     & $0.11$                         & $0.064$                      &  $0.847$    \\
L-shaped phantom     & $3.00$                         & $621$                     & $-81$                     & $-268$                    & $-0.16$                        & $0.072$                      &   $0.796$   \\
                     & $2.60$                         & $670$                     & $266$                     & $-8$                      & $0.16$                         & $0.083$                      &  $0.785$    \\
                     & $2.89$                         & $641$                     & $21$                      & $-192$                    & $-0.08$                        & $0.083$                      & $0.764$     \\
\hline
                     & $2.79$                         & $641$                     & $-134$                    & $-89$                     & $-0.02$                        & $0.081$                      &  $0.693$    \\
                     & $3.16$                         & $593$                     & $-223$                    & $-370$                    & $-0.34$                        & $0.065$                      &  $0.834$    \\
Phantom with a hole  & $2.90$                         & $638$                     & $8$                    & $-204$                    & $-0.04$                        & $0.122$                      &    $0.709$  \\
                     & $2.57$                         & $699$                     & $305$                     & $28$                      & $0.24$                         & $0.077$                      &  $0.789$    \\
                     & $2.58$                         & $716$                     & $291$                     & $22$                      & $0.27$                         & $0.045$                      &  $0.928$   \\
\hline
\end{tabular}
    \caption{\label{tab:diff_cal_disk_params} Relative error $\varepsilon_{\mathrm{rel}}$ and SSIM for full-angle FBP reconstructions of log phantom obtained with different parametrisations. The parametrisations were found using the L-shaped calibration phantom and the calibration phantom with a hole. True parameters are provided as a reference. Geometry parameter vectors for the L-shaped phantom correspond to those in Figure~\ref{fig:phantom_diff_parametrisations} }
    \end{table*}

    We simulated the scanning geometry, computed forward projections, Tikhonov and FBP reconstructions  using ODL. To compute MAP estimates we used the low-memory variant of the Broyden–Fletcher–Goldfarb–Shanno (BFGS) algorithm implemented in Julia.

    To assess the quality of reconstruction, we report relative reconstruction error defined as 

    \begin{equation*}
        \varepsilon_{\mathrm{rel}} = \frac{\| \hat{\mathbf{x}} - \mathbf{x}_{\mathrm{true}}\|_2}{\|\mathbf{x}_{\mathrm{true}}\|_2},
    \end{equation*}
    where image reconstruction $\hat{\mathbf{x}}$ and reference image  $\mathbf{x}_{\mathrm{true}}$ are in the vector form. As the other image quality metric, we employ the structural similarity index measure (SSIM) since it is shown to mimics well the perceived quality of an image by the human visual system \cite{WangBovik2004, WangBovik2009}. In numerical experiments, the full-angle FBP reconstruction is used as a reference.
    

    \subsection{Synthetic data analysis}\label{sec:synth_data}
    
    Here, we illustrate the proposed method for geometry calibration with an application to a digital log phantom (Figure~\ref{fig:log_phantom}). The phantom consists of concentric rings modelling tree growth rings, two ellipses representing knots inside a tree trunk, and one small ellipse with higher density representing a foreign object, for example, a metallic bullet.

    \begin{figure}[]
        \centering
        \includegraphics[width=1\linewidth]{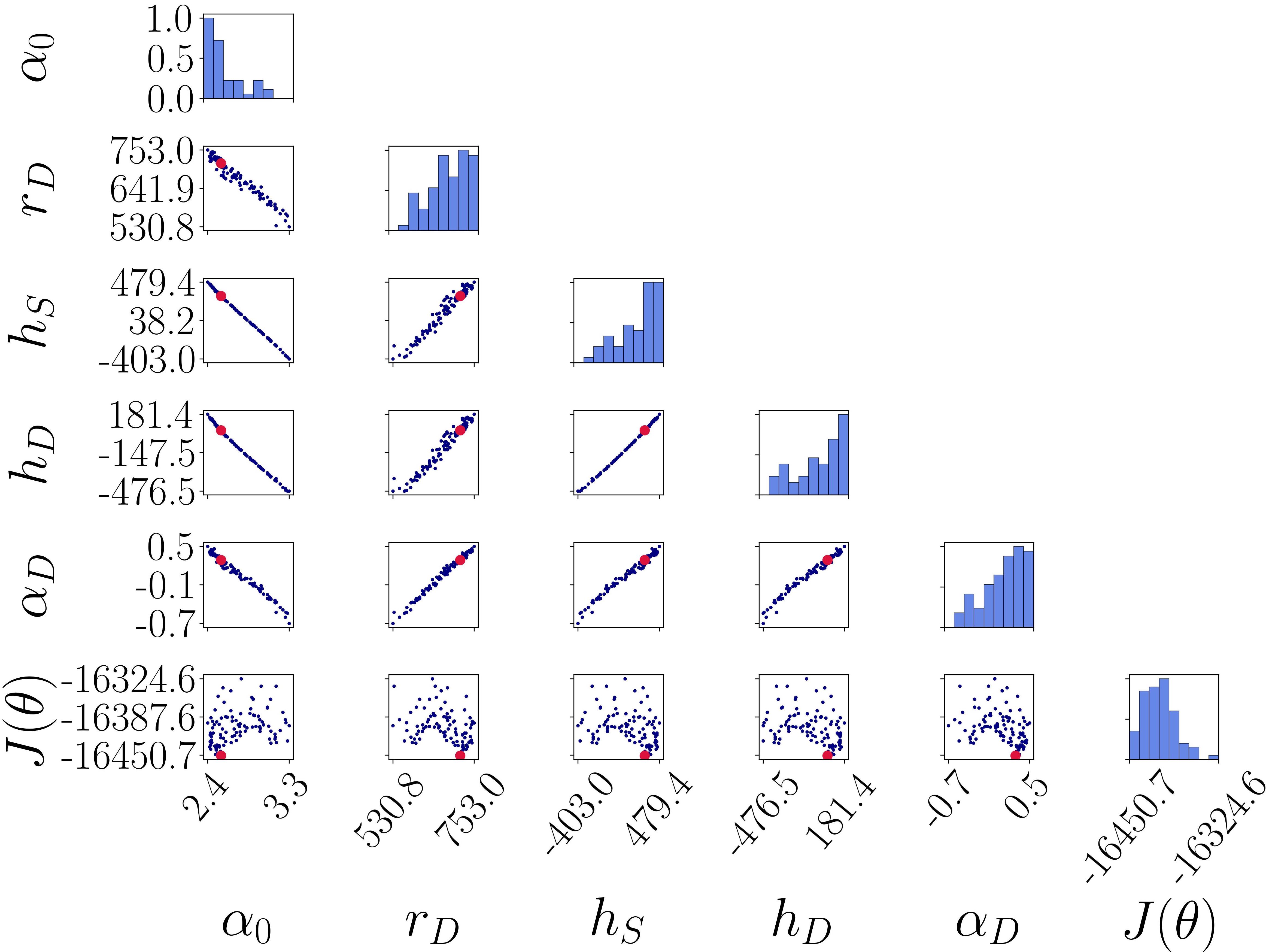}
        \caption{Pairwise plot of geometry parameters $\boldsymbol{\theta}=[\alpha_0, r_{D}, h_{S}, h_{D}, \alpha_D]^T$ and the objective function $J(\boldsymbol{\theta})$. Optimal geometry parameter values (in blue) were obtained by running the calibration method with \textit{full-angle} (360 projections) measurements of the L-shaped calibration phantom  100 times. True values of $\boldsymbol{\theta}$ and $J(\boldsymbol{\theta})$ are marked as red dots}\label{fig:params_L}
    \end{figure}  

    \begin{figure}[]
        \centering
        \includegraphics[width=1\linewidth]{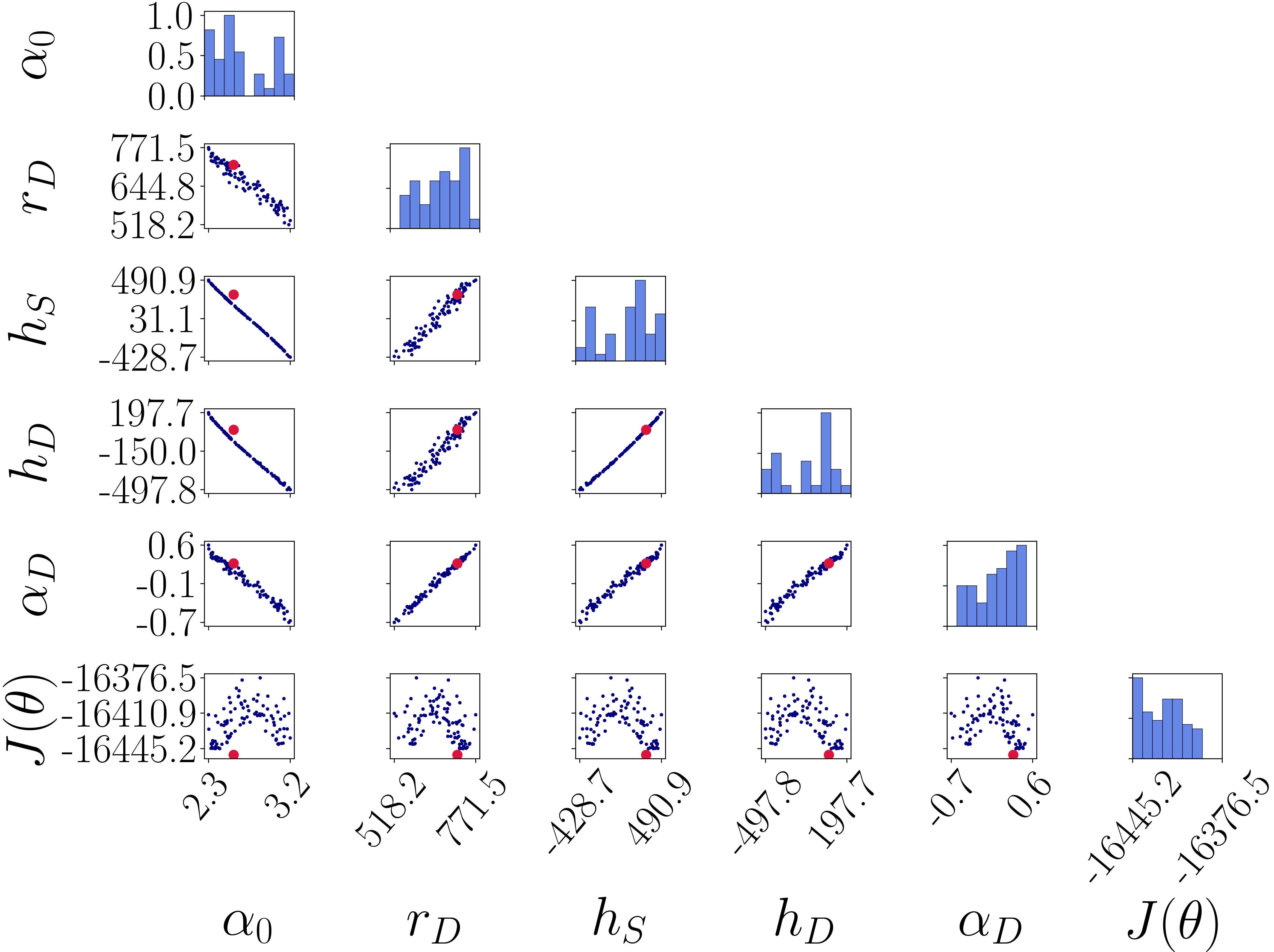}
        \caption{Pairwise plot of geometry parameters $\boldsymbol{\theta}=[\alpha_0, r_{D}, h_{S}, h_{D}, \alpha_D]^T$ and the objective function $J(\boldsymbol{\theta})$. Optimal geometry parameter values (in blue) were obtained by running the calibration method with \textit{sparse} (20 projections) measurements of the L-shaped calibration phantom  100 times. True values of $\boldsymbol{\theta}$ and $J(\boldsymbol{\theta})$ are marked as red dots}\label{fig:params_L_sparse}
    \end{figure}  

    \begin{figure}[]
        \centering
        \includegraphics[width=1\linewidth]{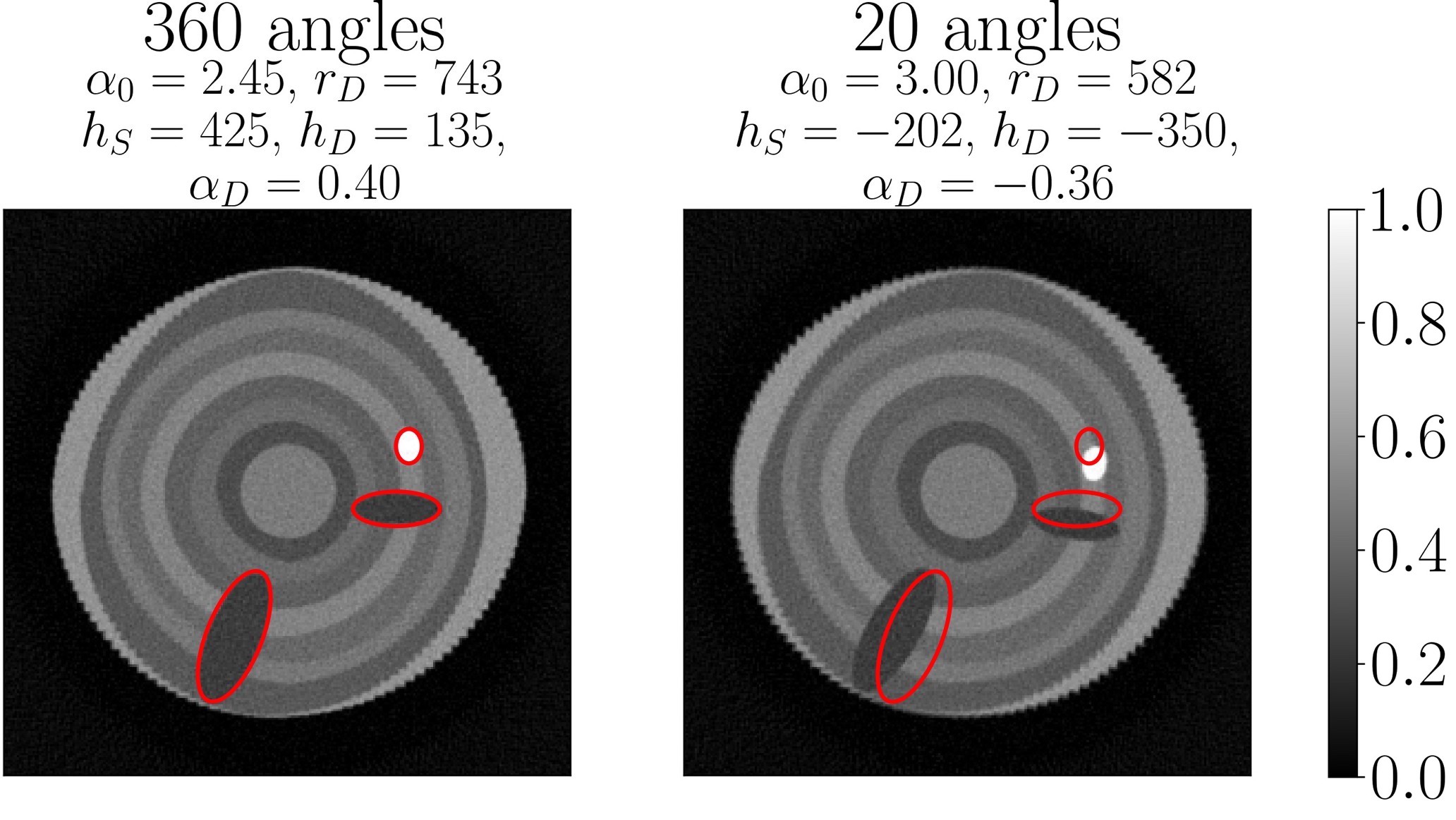}
        \caption{Full-angle reconstructions of log phantom. Left: geometry parameters are estimated using \textit{full-angle} measurements (360 projection angles); right: geometry parameters are estimated using \textit{sparse} measurements (20 projection angles)}\label{fig:phantom_compare}
    \end{figure}  
    
    We simulated full-angle tomography measurements of the log phantom and images of two calibration phantoms using ODL with true geometry parameter values $[\alpha_0, r_{D}, h_{S}, h_{D}, \alpha_D]^T = [2.55, 715, 320, 44, 0.28]^T$. To avoid the inverse crime \cite{Kaip2005}, we created the synthetic measurement data on a grid with $N = 1013$, then interpolated to a coarser grid with  $N = 256$ and added relative error noise of $2\%$. 

    \begin{figure*}[]
        \centering
        \includegraphics[width=1\linewidth]{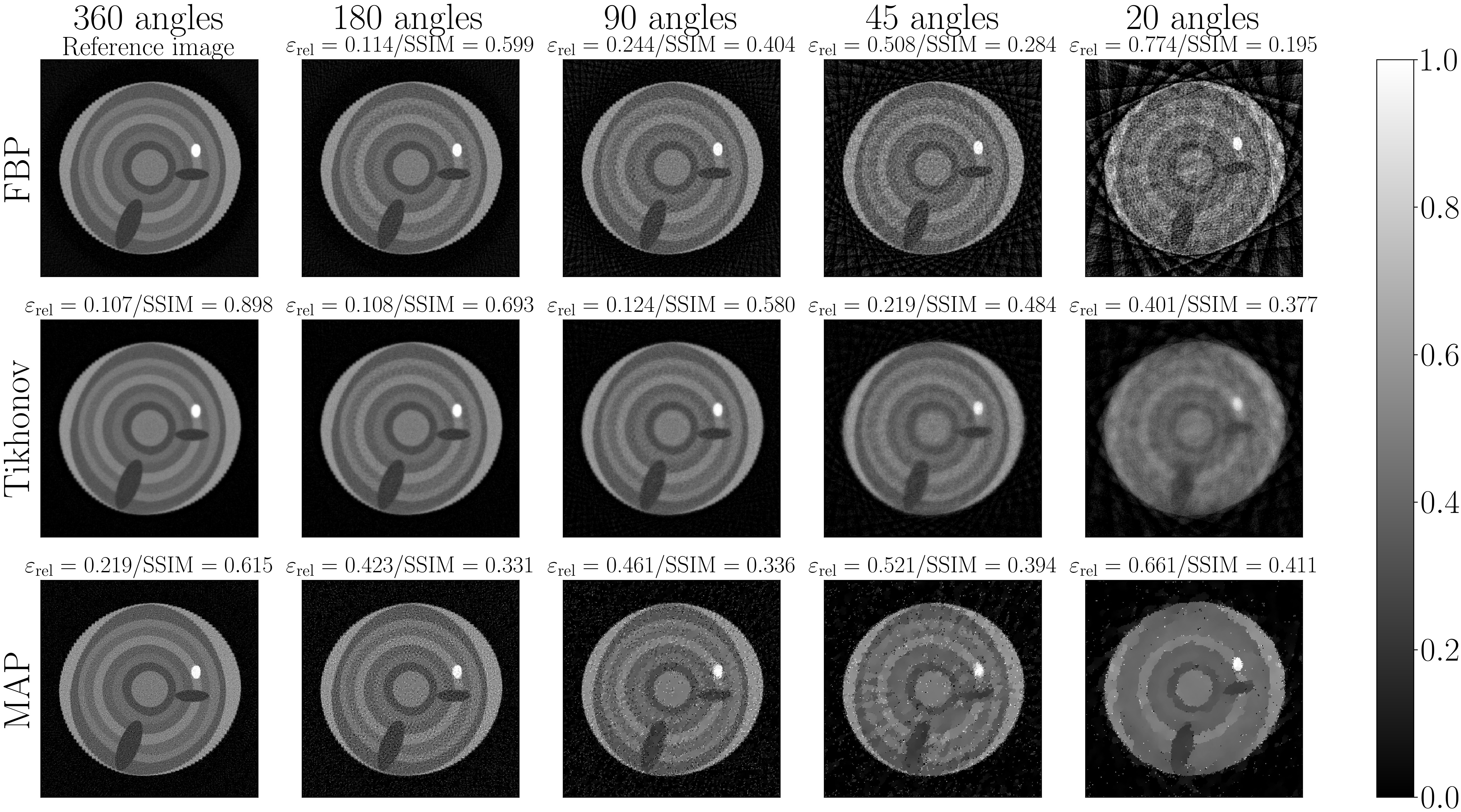}
        \caption{Reconstructions of the log phantom obtained by different reconstruction methods  with various numbers of projection angles (image resolution $256 \times 256$ pixels). The geometry parameters were obtained by the calibration method using the full-angle measurements of the L-shaped calibration phantom}\label{fig:phantom_recos}
    \end{figure*}  
    
    Figure~\ref{fig:ph_bad_recos} shows the effect of misspecified geometry parameters on the quality of full-angle FBP reconstructions of log phantom. We tested various combinations of geometry parameters (zero value means that the corresponding geometry parameter is misspecified, otherwise --- specified correctly):
    
    \begin{enumerate}[label=(\alph*)]
        \item $[\alpha_0, r_{D}, 0, 0, 0]^T$
        \item $[\alpha_0, r_{D}, h_{S}, 0, 0]^T$
        \item $[\alpha_0, r_{D}, h_{S}, h_{D}, 0]^T$
        \item $[\alpha_0, r_{D}, 0, 0, \alpha_D]^T$
        \item $[\alpha_0, r_{D} + \varepsilon, h_{S}, h_{D}, \alpha_D]^T$, where $\varepsilon$ is a small positive number (pertubation);
        \item $[\alpha_0, r_{D}, h_{S}, h_{D}, \alpha_D]^T$.
    \end{enumerate}
    In the figure, reconstructions corresponding to cases~(b), (c) and (e) contain severe artefacts such as duplicated features in the log phantom and halos, whereas reconstructions with misspecified parameter vectors~(a) and (d) are completely unidentifiable. 
    
    To tackle the issue of unknown geometry parameters, we ran the parameter search according to the method described in Section~\ref{sec:geometry}. In the parameter search, we aimed to find optimal values for five unknown geometry parameters, that is,  $\boldsymbol{\theta} = [\alpha_0, r_{D}, h_{S}, h_{D}, \alpha_D]^T$, based on simulated X-ray data of two calibration phantoms with true parameter values that we set to $[\alpha_0, r_{D}, h_{S}, h_{D}, \alpha_D]^T = [2.55, 715, 320, 44, 0.28]^T$.

    Figure~\ref{fig:phantom_diff_parametrisations} shows the full-angle FPB reconstructions with different geometry parameters found using the proposed calibration method. The geometry parameter search was done using the simulated full-angle measurements of the L-shaped calibration phantom, the reconstruction with the true geometry parameters is provided in the figure as a reference. It can be noted that all the different parametrisations provide satisfactory  quality in the FBP reconstructions although there is no unique optimal parameter set.

    We also ran the geometry parameter search using the other calibration phantom --- phantom with a hole. The obtained results were similar to those of L-shaped calibration phantom in Figure~\ref{fig:phantom_diff_parametrisations}, so we omit the plot. Instead,   in Table~\ref{tab:diff_cal_disk_params}, we report relative reconstruction error $\varepsilon_{\mathrm{rel}}$ and SSIM for FBP~reconstructions with different geometry parameters found by the geometry parameter search using two calibration phantoms. It can be concluded that both calibration phantoms perform well in the geometry parameter search algorithm and the~obtained estimates for the geometry parameters result in the reconstructions of good quality.

     \begin{figure*}[]
        \centering
        \includegraphics[width=1\linewidth]{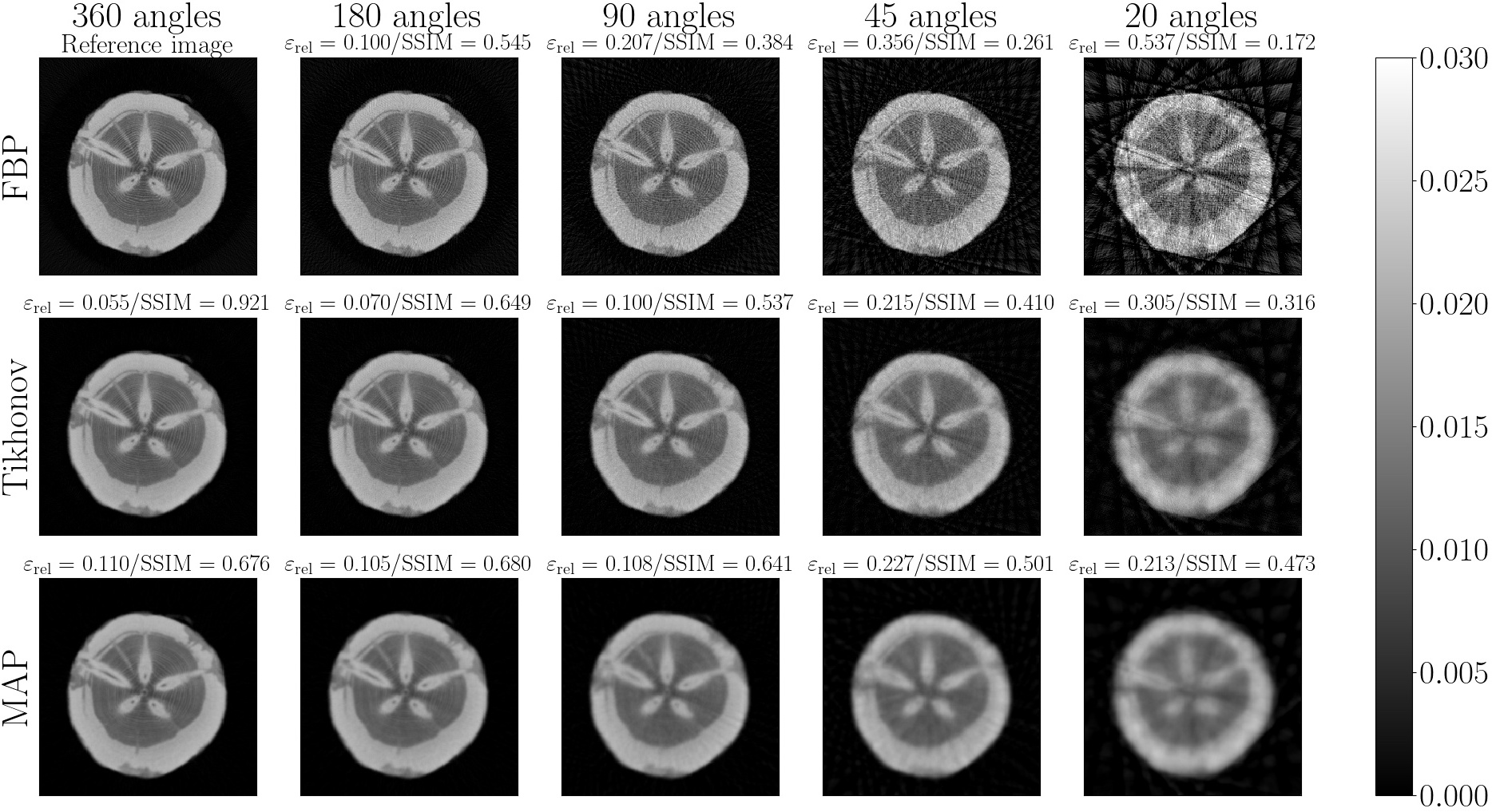}
        \caption{Two-dimensional reconstructions of the log slice obtained by different reconstruction methods with various numbers of projection angles (image resolution $256 \times 256$ pixels). The geometry parameters were obtained by the calibration method using the full-angle measurements of the L-shaped calibration phantom}\label{fig:recos}
    \end{figure*}    

    To provide a view on the uncertainty of the found parameters, we ran the optimisation algorithm with L-shaped phantom 100 times and created the pairwise plot of optimal parameters together with the true parameter values (see Figure~\ref{fig:params_L}). It can be noted that the parameters are highly correlated and their distribution is very narrow. When using the calibration phantom with a hole we observed similar behaviour of the parameters so we omitted the corresponding plot. 
    
    In addition, we tested the performance of the calibration method when the sparse tomography measurements of the calibration phantoms were used. In Figure~\ref{fig:params_L_sparse}, we show the results of geometry parameter search using sparse measurements (20 projection angles) of L-shaped phantom. The results are similar to those in experiment in Figure~\ref{fig:params_L}, where the full-angle measurements of the calibration phantom were used. In Figure~\ref{fig:phantom_compare}, two full-angle FBP reconstructions of log phantom are shown for comparison: geometry parameters for the left reconstruction were obtained using the full-angle measurements of L-shaped phantom while for the right reconstruction, sparse measurements of the phantom were used. Red contour lines highlighting the phantom features are used to show that there is a small angular misalignment between two reconstructions. 
    Despite the misalignment both reconstructions are of good quality, therefore, the proposed calibration method works in the sparse-angle scenario as well. 
    
    Although the scanning geometry is estimated correctly using the calibration method, severe artefacts still appear in the reconstructions if classical reconstruction approaches such as FBP and Tikhonov regularisation are used and the number of projection angles is very low. To tackle the issue of artefacts, we employed Bayesian inversion with
    edge-preserving Cauchy priors from Section~\ref{sec:reco_methods}. As shown in Figure~\ref{fig:phantom_recos}, in the sparse case (20 projection angles), the FBP reconstruction contains severe streaking artefacts and the Tikhonov regularised reconstruction is extremely blurry, while the quality of the MAP estimate with a Cauchy prior is still good enough and allows for detection of knots and other features in the log phantom. 

    For MAP reconstructions in the figure, the relative error $\varepsilon_\mathrm{rel}$ grows representing the expected quality deterioration as the number of projection angles decreases. However, SSIM rises suggesting the quality improvement, which contradicts the visual quality assessment and $\varepsilon_\mathrm{rel}$. This phenomena can be explained by the reduced ability of SSIM to represent the visual quality for CT reconstructions~\cite{PamNou2015} and the images containing salt-and-pepper (impulse) noise~\cite{LinChen2022}.


    \subsection{Real data analysis}

    In this section, the proposed method for geometry calibration is applied to real X-ray data of a log. The X-ray measurements  were obtained with an X-ray log scanning system provided by Finnos~Oy. A tree log was scanned jointly with the wooden calibration phantoms (Figure~\ref{fig:calibration_phantoms}) attached to the saw cut of the one side of the log. The calibration phantom plane was, therefore, perpendicular to the log axis of rotation.
    
    As a result of the sequential (``step and shoot'') X-ray scanning, we obtained projection images capturing the log with attached calibration object from 360 different angles. The  width of the projection image is fixed and corresponds to the number of detector elements $n_{D}$. The  height of the projection image depends on the number of slices that have been scanned (the width of 1 slice is $1$ px, or equivalently, $2$ mm), that is, the length of the log scanned. Due to the peculiarities of the measurement system, there is an issue of missing pixel rows at either the top or bottom of the projection image, and  there is a need to overlay images corresponding to projections of the same log taken from different angles using an image registration technique \cite{Zit2003}. In the context of our task, we adhere to the following image registration method consisting of four consecutive steps: feature detection, feature matching, transform model estimation, image resampling and transformation. 
    The algorithm was implemented using the OpenCV Python library.

    \begin{figure}[]
        \centering
        \includegraphics[width=1\linewidth]{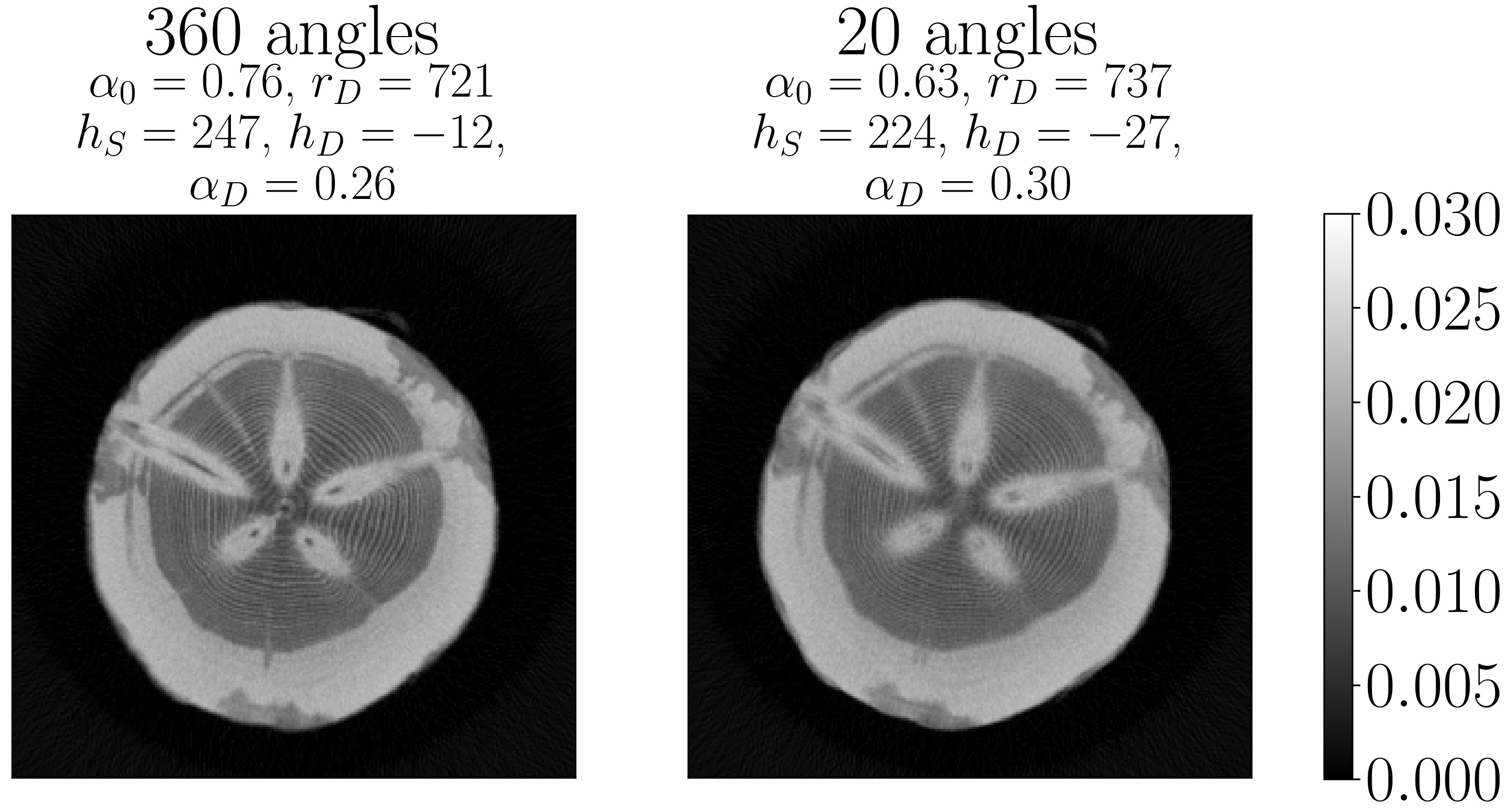}
        \caption{Full-angle reconstructions of log slice. Left: geometry parameters are estimated using \textit{full-angle} measurements (360 projection angles); right: geometry parameters are estimated using \textit{sparse} measurements (20 projection angles)}\label{fig:real_data_compare}
    \end{figure} 
    
    We estimated five unknown geometry parameters (the first scanning angle, distance between the COR and detector, the source shift, the detector shift, and the detector tilt) using the reference images of the calibration phantoms and the corresponding X-ray data obtained from the measurement device. In Figure~\ref{fig:recos}, we show the cross-sectional reconstructions of one slice of the log obtained after geometry parameter estimation using the full-angle measurements of the L-shaped calibration phantom. The figure compares different reconstruction methods (FBP, Tikhonov regularisation, MAP estimate) for various numbers of projection angles. In sparse measurements scenario (45 and 20 projection angles), the quality of FBP reconstructions is compromised by streaking artefacts, while Tikhonov and MAP reconstructions show relatively better quality.
    
    To show the strength of the calibration method applied to sparse data, we provide two full-angle reconstructions of the same cross-sectional slice of the log obtained with different parametrisations (see Figure~\ref{fig:real_data_compare}).  Geometry parameters of the left reconstruction were obtained using the full-angle measurements of the the L-shaped calibration phantom, whereas parameters of the right reconstruction were estimated from sparse measurements (20 projection angles). Both reconstructions are of good quality and can be used to locate the tree knots and possible defects in the log. 
 
    \section{Conclusions}
    \label{sec:discussion}
    
    In this paper, we have proposed an algorithm for the estimation of unknown geometry parameters in fan-beam X-ray computed tomography (CT) for log imaging. 
    The algorithm employs  easy-to-produce calibration phantoms that are scanned jointly with the log and applies differential evolution optimisation to the objective function represented by the maximum cross-correlation between the ground truth image of the calibration phantom and its filtered backprojection (FBP) reconstruction. The algorithm is able to estimate geometry parameters not only from full-angle measurements of the calibration phantom but also from sparse measurements.

    We demonstrated numerically that there might be multiple parameter vectors that give almost the same cross-correlation with the reference object when used in the FBP reconstruction, that is, different sets of optimal parameters deliver an adequate quality of reconstructions. 
    However,  the uniqueness of the geometry parameters of the measurement equipment does not pose any interest, since we are only interested in the quality of the reconstructions.

    Additionally, we demonstrated the effect of misspecified geometry parameters, including the first scanning angle, the centre of rotation (COR) to detector distance, the source and detector shifts, and the detector tilt, on the quality of CT reconstructions. 
    For reconstruction purposes, we employed Bayesian inversion (the MAP estimate of the posterior distribution with Cauchy difference priors) as opposed to classical reconstruction approaches such as FBP and Tikhonov regularisation in order to improve the reconstruction quality in the sparse CT setting. Bayesian inversion showed better performance on sparse data than FBP method that suffered from severe streaking artefacts compromising the quality of reconstructions. 
    

    \section*{Acknowledgments}
    \label{sec:acknowledgments}
    
    This work was funded by the Academy of Finland (project numbers 336787 and 334816). 
    
    \section*{Data availability}
    \label{sec:data}
    All the codes needed to generate the synthetic data and reconstructions are available on GitHub: \url{https://github.com/AngelinaSen/geometry_parameter_estimation}.  
    The real log tomography datasets analysed during the current study are not made public but are available from the corresponding author on reasonable request.
    
    \bibliography{sn-article.bib}    


\begin{thebibliography}{40}
\ifx \bisbn   \undefined \def \bisbn  #1{ISBN #1}\fi
\ifx \binits  \undefined \def \binits#1{#1}\fi
\ifx \bauthor  \undefined \def \bauthor#1{#1}\fi
\ifx \batitle  \undefined \def \batitle#1{#1}\fi
\ifx \bjtitle  \undefined \def \bjtitle#1{#1}\fi
\ifx \bvolume  \undefined \def \bvolume#1{\textbf{#1}}\fi
\ifx \byear  \undefined \def \byear#1{#1}\fi
\ifx \bissue  \undefined \def \bissue#1{#1}\fi
\ifx \bfpage  \undefined \def \bfpage#1{#1}\fi
\ifx \blpage  \undefined \def \blpage #1{#1}\fi
\ifx \burl  \undefined \def \burl#1{\textsf{#1}}\fi
\ifx \doiurl  \undefined \def \doiurl#1{\url{https://doi.org/#1}}\fi
\ifx \betal  \undefined \def \betal{\textit{et al.}}\fi
\ifx \binstitute  \undefined \def \binstitute#1{#1}\fi
\ifx \binstitutionaled  \undefined \def \binstitutionaled#1{#1}\fi
\ifx \bctitle  \undefined \def \bctitle#1{#1}\fi
\ifx \beditor  \undefined \def \beditor#1{#1}\fi
\ifx \bpublisher  \undefined \def \bpublisher#1{#1}\fi
\ifx \bbtitle  \undefined \def \bbtitle#1{#1}\fi
\ifx \bedition  \undefined \def \bedition#1{#1}\fi
\ifx \bseriesno  \undefined \def \bseriesno#1{#1}\fi
\ifx \blocation  \undefined \def \blocation#1{#1}\fi
\ifx \bsertitle  \undefined \def \bsertitle#1{#1}\fi
\ifx \bsnm \undefined \def \bsnm#1{#1}\fi
\ifx \bsuffix \undefined \def \bsuffix#1{#1}\fi
\ifx \bparticle \undefined \def \bparticle#1{#1}\fi
\ifx \barticle \undefined \def \barticle#1{#1}\fi
\bibcommenthead
\ifx \bconfdate \undefined \def \bconfdate #1{#1}\fi
\ifx \botherref \undefined \def \botherref #1{#1}\fi
\ifx \url \undefined \def \url#1{\textsf{#1}}\fi
\ifx \bchapter \undefined \def \bchapter#1{#1}\fi
\ifx \bbook \undefined \def \bbook#1{#1}\fi
\ifx \bcomment \undefined \def \bcomment#1{#1}\fi
\ifx \oauthor \undefined \def \oauthor#1{#1}\fi
\ifx \citeauthoryear \undefined \def \citeauthoryear#1{#1}\fi
\ifx \endbibitem  \undefined \def \endbibitem {}\fi
\ifx \bconflocation  \undefined \def \bconflocation#1{#1}\fi
\ifx \arxivurl  \undefined \def \arxivurl#1{\textsf{#1}}\fi
\csname PreBibitemsHook\endcsname

\bibitem{Zol2019}
\begin{bchapter}
\bauthor{\bsnm{Zolotarev}, \binits{F.}},
\bauthor{\bsnm{Eerola}, \binits{T.}},
\bauthor{\bsnm{Lensu}, \binits{L.}},
\bauthor{\bsnm{K{\"a}lvi{\"a}inen}, \binits{H.}},
\bauthor{\bsnm{Haario}, \binits{H.}},
\bauthor{\bsnm{Heikkinen}, \binits{J.}},
\bauthor{\bsnm{Kauppi}, \binits{T.}}:
\bctitle{Timber tracing with multimodal encoder-decoder networks}.
In: \bbtitle{International Conference on Computer Analysis of Images and Patterns},
pp. \bfpage{342}--\blpage{353}
(\byear{2019})
\end{bchapter}
\endbibitem

\bibitem{Flod2008}
\begin{barticle}
\bauthor{\bsnm{Flodin}, \binits{J.}},
\bauthor{\bsnm{Oja}, \binits{J.}},
\bauthor{\bsnm{Grönlund}, \binits{A.}}:
\batitle{Fingerprint traceability of sawn products using x-ray log scanning and sawn timber surface scanning}.
\bjtitle{Forest products journal}
\bvolume{58},
\bfpage{100}--\blpage{105}
(\byear{2008})
\end{barticle}
\endbibitem

\bibitem{Tar2004}
\begin{bbook}
\bauthor{\bsnm{Tarantola}, \binits{A.}}:
\bbtitle{Inverse Problem Theory and Methods for Model Parameter Estimation}.
\bpublisher{Society for Industrial and Applied Mathematics},
\blocation{Philadelphia}
(\byear{2004})
\end{bbook}
\endbibitem

\bibitem{Kaip2005}
\begin{bbook}
\bauthor{\bsnm{Kaipio}, \binits{J.}},
\bauthor{\bsnm{Somersalo}, \binits{E.}}:
\bbtitle{Statistical and Computational Inverse Problems}.
\bpublisher{Springer},
\blocation{Dordrecht}
(\byear{2005})
\end{bbook}
\endbibitem

\bibitem{Stu2010}
\begin{barticle}
\bauthor{\bsnm{Stuart}, \binits{A.M.}}:
\batitle{Inverse problems: A {B}ayesian perspective}.
\bjtitle{Acta Numerica}
\bvolume{19},
\bfpage{451}--\blpage{559}
(\byear{2010}).
\doiurl{10.1017/S0962492910000061}
\end{barticle}
\endbibitem

\bibitem{Nat1986}
\begin{bbook}
\bauthor{\bsnm{Natterer}, \binits{F.}}:
\bbtitle{The Mathematics of Computerized Tomography}.
\bpublisher{Wiley},
\blocation{Chicago}
(\byear{1986})
\end{bbook}
\endbibitem

\bibitem{Suur2020}
\begin{botherref}
\oauthor{\bsnm{Suuronen}, \binits{J.}},
\oauthor{\bsnm{Emzir}, \binits{M.}},
\oauthor{\bsnm{Lasanen}, \binits{S.}},
\oauthor{\bsnm{S{\"a}rkk{\"a}}, \binits{S.}},
\oauthor{\bsnm{Roininen}, \binits{L.}}:
Enhancing industrial {X}-ray tomography by data-centric statistical methods.
Data-Centric Engineering
\textbf{1}
(2020)
\end{botherref}
\endbibitem

\bibitem{LiGull1994}
\begin{barticle}
\bauthor{\bsnm{Li}, \binits{J.}},
\bauthor{\bsnm{Jaszczak}, \binits{R.}},
\bauthor{\bsnm{Wang}, \binits{H.}},
\bauthor{\bsnm{Gullberg}, \binits{G.}},
\bauthor{\bsnm{Greer}, \binits{K.}},
\bauthor{\bsnm{Coleman}, \binits{E.}}:
\batitle{A cone beam {SPECT} reconstruction algorithm with a displaced center of rotation}.
\bjtitle{Medical physics}
\bvolume{21},
\bfpage{145}--\blpage{52}
(\byear{1994})
\end{barticle}
\endbibitem

\bibitem{Huili1998}
\begin{barticle}
\bauthor{\bsnm{Wang}, \binits{H.}},
\bauthor{\bsnm{Smith.}, \binits{M.F.}},
\bauthor{\bsnm{Stone}, \binits{C.D.}},
\bauthor{\bsnm{Jaszczak}, \binits{R.J.}}:
\batitle{Astigmatic single photon emission computed tomography imaging with a displaced center of rotation}.
\bjtitle{Medical physics}
\bvolume{25},
\bfpage{1493}--\blpage{1501}
(\byear{1998})
\end{barticle}
\endbibitem

\bibitem{Jereb2012}
\begin{bchapter}
\bauthor{\bsnm{Dennerlein}, \binits{F.}},
\bauthor{\bsnm{Jerebko}, \binits{A.}}:
\bctitle{Geometric jitter compensation in cone-beam {CT} through registration of directly and indirectly filtered projections}.
In: \bbtitle{Nuclear Science Symposium and Medical Imaging Conference (NSS/MIC)},
pp. \bfpage{2892}--\blpage{2895}
(\byear{2012}).
\doiurl{10.1109/NSSMIC.2012.6551660}
\end{bchapter}
\endbibitem

\bibitem{Ferr2015}
\begin{botherref}
\oauthor{\bsnm{Ferrucci}, \binits{M.}},
\oauthor{\bsnm{Leach}, \binits{R.K.}},
\oauthor{\bsnm{Giusca}, \binits{C.}},
\oauthor{\bsnm{Carmignato}, \binits{S.}},
\oauthor{\bsnm{Dewulf}, \binits{W.}}:
Towards geometrical calibration of {X}-ray computed tomography systems --- {A}~review
\textbf{26}(9),
092003
(2015).
\doiurl{10.1088/0957-0233/26/9/092003}
\end{botherref}
\endbibitem

\bibitem{Zem2023}
\begin{barticle}
\bauthor{\bsnm{Zemek}, \binits{M.}},
\bauthor{\bsnm{Šalplachta}, \binits{J.}},
\bauthor{\bsnm{Zikmund}, \binits{T.}},
\bauthor{\bsnm{Omote}, \binits{K.}},
\bauthor{\bsnm{Takeda}, \binits{Y.}},
\bauthor{\bsnm{Oberta}, \binits{P.}},
\bauthor{\bsnm{Kaiser}, \binits{J.}}:
\batitle{Automatic marker-free estimation methods for the axis of rotation in sub-micron {X}-ray computed tomography}.
\bjtitle{Tomography of Materials and Structures}
\bvolume{1},
\bfpage{100002}
(\byear{2023}).
\doiurl{10.1016/j.tmater.2022.100002}
\end{barticle}
\endbibitem

\bibitem{Gull1987}
\begin{botherref}
\oauthor{\bsnm{Gullberg}, \binits{G.T.}},
\oauthor{\bsnm{Tsui}, \binits{B.M.W.}},
\oauthor{\bsnm{Crawford}, \binits{C.R.}},
\oauthor{\bsnm{Edgerton}, \binits{E.R.}}:
Estimation of geometrical parameters for fan beam tomography
\textbf{32}(12),
1581--1594
(1987)
\end{botherref}
\endbibitem

\bibitem{Cho2005}
\begin{barticle}
\bauthor{\bsnm{Cho}, \binits{Y.}},
\bauthor{\bsnm{Moseley}, \binits{D.}},
\bauthor{\bsnm{Siewerdsen}, \binits{J.}},
\bauthor{\bsnm{Jaffray}, \binits{D.}}:
\batitle{Accurate technique for complete geometric calibration of cone-beam computed tomography systems}.
\bjtitle{Medical physics}
\bvolume{32},
\bfpage{968}--\blpage{83}
(\byear{2005}).
\doiurl{10.1118/1.1869652}
\end{barticle}
\endbibitem

\bibitem{Ouadah2016}
\begin{barticle}
\bauthor{\bsnm{Ouadah}, \binits{S.}},
\bauthor{\bsnm{Stayman}, \binits{J.}},
\bauthor{\bsnm{Gang}, \binits{G.}},
\bauthor{\bsnm{Ehtiati}, \binits{T.}},
\bauthor{\bsnm{Siewerdsen}, \binits{J.}}:
\batitle{Self-calibration of cone-beam {CT} geometry using 3{D}–2{D} image registration}.
\bjtitle{Physics in Medicine and Biology}
\bvolume{61},
\bfpage{2613}--\blpage{2632}
(\byear{2016}).
\doiurl{10.1088/0031-9155/61/7/2613}
\end{barticle}
\endbibitem

\bibitem{Uribe2021}
\begin{barticle}
\bauthor{\bsnm{Uribe}, \binits{F.}},
\bauthor{\bsnm{Bardsley}, \binits{J.M.}},
\bauthor{\bsnm{Dong}, \binits{Y.}},
\bauthor{\bsnm{Hansen}, \binits{P.C.}},
\bauthor{\bsnm{Riis}, \binits{N.A.B.}}:
\batitle{A hybrid {G}ibbs sampler for edge-preserving tomographic reconstruction with uncertain view angles}.
\bjtitle{SIAM/ASA Journal on Uncertainty Quantification}
\bvolume{10},
\bfpage{1293}--\blpage{1320}
(\byear{2021})
\end{barticle}
\endbibitem

\bibitem{Hansen2021}
\begin{barticle}
\bauthor{\bsnm{Riis}, \binits{N.}},
\bauthor{\bsnm{Dong}, \binits{Y.}},
\bauthor{\bsnm{Hansen}, \binits{P.C.}}:
\batitle{Computed tomography reconstruction with uncertain view angles by iteratively updated model discrepancy}.
\bjtitle{Journal of Mathematical Imaging and Vision}
\bvolume{63},
\bfpage{133}--\blpage{143}
(\byear{2021}).
\doiurl{10.1007/s10851-020-00972-7}
\end{barticle}
\endbibitem

\bibitem{Ped2023}
\begin{barticle}
\bauthor{\bsnm{Pedersen}, \binits{F.H.}},
\bauthor{\bsnm{Jørgensen}, \binits{J.S.}},
\bauthor{\bsnm{Andersen}, \binits{M.S.}}:
\batitle{A bayesian approach to {CT} reconstruction with uncertain geometry}.
\bjtitle{Applied Mathematics in Science and Engineering}
\bvolume{31}(\bissue{1}),
\bfpage{2166041}
(\byear{2023}).
\doiurl{10.1080/27690911.2023.2166041}
\end{barticle}
\endbibitem

\bibitem{Gen2021}
\begin{botherref}
\oauthor{\bsnm{Genzel}, \binits{M.}},
\oauthor{\bsnm{Macdonald}, \binits{J.}},
\oauthor{\bsnm{M{\"a}rz}, \binits{M.}}:
AAPM DL-Sparse-View CT Challenge submission report: Designing an iterative network for fanbeam-CT with unknown geometry.
arXiv
(2021).
\url{https://arxiv.org/abs/2106.00280}
\end{botherref}
\endbibitem

\bibitem{Xie2020}
\begin{botherref}
\oauthor{\bsnm{Xie}, \binits{M.}},
\oauthor{\bsnm{Sun}, \binits{Y.}},
\oauthor{\bsnm{Liu}, \binits{J.}},
\oauthor{\bsnm{Wohlberg}, \binits{B.}},
\oauthor{\bsnm{Kamilov}, \binits{U.S.}}:
Joint reconstruction and calibration using regularization by denoising.
arXiv
(2020).
\url{https://arxiv.org/abs/2011.13391}
\end{botherref}
\endbibitem

\bibitem{Suur2022}
\begin{botherref}
\oauthor{\bsnm{Suuronen}, \binits{J.}},
\oauthor{\bsnm{Chada}, \binits{N.}},
\oauthor{\bsnm{Roininen}, \binits{L.}}:
Cauchy {M}arkov random field priors for {B}ayesian inversion.
Statistics and Computing
\textbf{32}(2)
(2022)
\end{botherref}
\endbibitem

\bibitem{Rad1986}
\begin{barticle}
\bauthor{\bsnm{Radon}, \binits{J.}}:
\batitle{On the determination of functions from their integral values along certain manifolds}.
\bjtitle{IEEE Transactions on Medical Imaging}
\bvolume{5}(\bissue{4}),
\bfpage{170}--\blpage{176}
(\byear{1986})
\end{barticle}
\endbibitem

\bibitem{Kak2001}
\begin{bbook}
\bauthor{\bsnm{Kak}, \binits{A.C.}},
\bauthor{\bsnm{Slaney}, \binits{M.}}:
\bbtitle{Principles of Computerized Tomographic Imaging}.
\bpublisher{Society of Industrial and Applied Mathematics},
\blocation{Philadelphia}
(\byear{2001})
\end{bbook}
\endbibitem

\bibitem{Yu2002}
\begin{barticle}
\bauthor{\bsnm{Yu}, \binits{D.}},
\bauthor{\bsnm{Fessler}, \binits{J.}}:
\batitle{Edge-preserving tomographic reconstruction with nonlocal regularization}.
\bjtitle{Medical Imaging, IEEE Transactions on}
\bvolume{21},
\bfpage{159}--\blpage{173}
(\byear{2002}).
\doiurl{10.1109/42.993134}
\end{barticle}
\endbibitem

\bibitem{Xu2019}
\begin{barticle}
\bauthor{\bsnm{Xu}, \binits{J.}},
\bauthor{\bsnm{Zhao}, \binits{Y.}},
\bauthor{\bsnm{Li}, \binits{H.}},
\bauthor{\bsnm{Zhang}, \binits{P.}}:
\batitle{An image reconstruction model regularized by edge-preserving diffusion and smoothing for limited-angle computed tomography}.
\bjtitle{Inverse Problems}
\bvolume{35},
\bfpage{085004}
(\byear{2019}).
\doiurl{10.1088/1361-6420/ab08f9}
\end{barticle}
\endbibitem

\bibitem{Hama2013}
\begin{barticle}
\bauthor{\bsnm{H{\"a}m{\"a}l{\"a}inen}, \binits{K.}},
\bauthor{\bsnm{Kallonen}, \binits{A.}},
\bauthor{\bsnm{Kolehmainen}, \binits{V.}},
\bauthor{\bsnm{Lassas}, \binits{M.}},
\bauthor{\bsnm{Niinim{\"a}ki}, \binits{K.}},
\bauthor{\bsnm{Siltanen}, \binits{S.}}:
\batitle{Sparse tomography}.
\bjtitle{SIAM Journal on Scientific Computing}
\bvolume{35},
\bfpage{644}--\blpage{665}
(\byear{2013}).
\doiurl{10.1137/120876277}
\end{barticle}
\endbibitem

\bibitem{Markkanen19}
\begin{botherref}
\oauthor{\bsnm{Markkanen}, \binits{M.}},
\oauthor{\bsnm{Roininen}, \binits{L.}},
\oauthor{\bsnm{Huttunen}, \binits{J.}},
\oauthor{\bsnm{Lasanen}, \binits{S.}}:
Cauchy difference priors for edge-preserving {B}ayesian inversion.
Journal of Inverse and Ill-posed Problems
\textbf{27}
(2019).
\doiurl{10.1515/jiip-2017-0048}
\end{botherref}
\endbibitem

\bibitem{StornPrice2020}
\begin{barticle}
\bauthor{\bsnm{Storn}, \binits{R.}},
\bauthor{\bsnm{Price}, \binits{K.}}:
\batitle{Differential evolution --- {A} simple and efficient heuristic for global optimization over continuous spaces}.
\bjtitle{Journal of Global Optimization}
\bvolume{11},
\bfpage{341}--\blpage{359}
(\byear{1997})
\end{barticle}
\endbibitem

\bibitem{Red2007}
\begin{barticle}
\bauthor{\bsnm{Reddy}, \binits{M.J.}},
\bauthor{\bsnm{Kumar}, \binits{D.N.}}:
\batitle{Multiobjective differential evolution with application to reservoir system optimization}.
\bjtitle{Journal of Computing in Civil Engineering}
\bvolume{21},
\bfpage{136}--\blpage{146}
(\byear{2007}).
\doiurl{10.1061/(ASCE)0887-3801(2007)21:2(136)}
\end{barticle}
\endbibitem

\bibitem{Leon2014}
\begin{bchapter}
\bauthor{\bsnm{Leon}, \binits{M.}},
\bauthor{\bsnm{Xiong}, \binits{N.}}:
\bctitle{Investigation of mutation strategies in differential evolution for solving global optimization problems},
vol. \bseriesno{8467}
(\byear{2014}).
\doiurl{10.1007/978-3-319-07173-2_32}
\end{bchapter}
\endbibitem

\bibitem{SciPy2020}
\begin{barticle}
\bauthor{\bsnm{Virtanen}, \binits{P.}},
\bauthor{\bsnm{Gommers}, \binits{R.}},
\bauthor{\bsnm{Oliphant}, \binits{T.E.}},
\bauthor{\bsnm{Haberland}, \binits{M.}},
\bauthor{\bsnm{Reddy}, \binits{T.}},
\bauthor{\bsnm{Cournapeau}, \binits{D.}},
\bauthor{\bsnm{Burovski}, \binits{E.}},
\bauthor{\bsnm{Peterson}, \binits{P.}},
\bauthor{\bsnm{Weckesser}, \binits{W.}},
\bauthor{\bsnm{Bright}, \binits{J.}},
\bauthor{\bsnm{{van der Walt}}, \binits{S.J.}},
\bauthor{\bsnm{Brett}, \binits{M.}},
\bauthor{\bsnm{Wilson}, \binits{J.}},
\bauthor{\bsnm{Millman}, \binits{K.J.}},
\bauthor{\bsnm{Mayorov}, \binits{N.}},
\bauthor{\bsnm{Nelson}, \binits{A.R.J.}},
\bauthor{\bsnm{Jones}, \binits{E.}},
\bauthor{\bsnm{Kern}, \binits{R.}},
\bauthor{\bsnm{Larson}, \binits{E.}},
\bauthor{\bsnm{Carey}, \binits{C.J.}},
\bauthor{\bsnm{Polat}, \binits{{\. I}.}},
\bauthor{\bsnm{Feng}, \binits{Y.}},
\bauthor{\bsnm{Moore}, \binits{E.W.}},
\bauthor{\bsnm{{VanderPlas}}, \binits{J.}},
\bauthor{\bsnm{Laxalde}, \binits{D.}},
\bauthor{\bsnm{Perktold}, \binits{J.}},
\bauthor{\bsnm{Cimrman}, \binits{R.}},
\bauthor{\bsnm{Henriksen}, \binits{I.}},
\bauthor{\bsnm{Quintero}, \binits{E.A.}},
\bauthor{\bsnm{Harris}, \binits{C.R.}},
\bauthor{\bsnm{Archibald}, \binits{A.M.}},
\bauthor{\bsnm{Ribeiro}, \binits{A.H.}},
\bauthor{\bsnm{Pedregosa}, \binits{F.}},
\bauthor{\bsnm{{van Mulbregt}}, \binits{P.}},
\bauthor{\bsnm{{SciPy 1.0 Contributors}}}:
\batitle{{{SciPy} 1.0: Fundamental Algorithms for Scientific Computing in Python}}.
\bjtitle{Nature Methods}
\bvolume{17},
\bfpage{261}--\blpage{272}
(\byear{2020}).
\doiurl{10.1038/s41592-019-0686-2}
\end{barticle}
\endbibitem

\bibitem{Nelder1965}
\begin{barticle}
\bauthor{\bsnm{Nelder}, \binits{J.A.}},
\bauthor{\bsnm{Mead}, \binits{R.}}:
\batitle{A simplex method for function minimization}.
\bjtitle{Computer Journal}
\bvolume{7},
\bfpage{308}--\blpage{313}
(\byear{1965})
\end{barticle}
\endbibitem

\bibitem{SA1989}
\begin{bbook}
\bauthor{\bsnm{Aarts}, \binits{E.}},
\bauthor{\bsnm{Korst}, \binits{J.}}:
\bbtitle{Simulated Annealing and Boltzmann Machines: A Stochastic Approach to Combinatorial Optimization and Neural Computing}.
\bpublisher{John Wiley \& Sons, Inc.},
\blocation{New York}
(\byear{1989})
\end{bbook}
\endbibitem

\bibitem{Horst2013}
\begin{bbook}
\bauthor{\bsnm{Horst}, \binits{R.}},
\bauthor{\bsnm{Pardalos}, \binits{P.M.}}:
\bbtitle{Handbook of Global Optimization}
vol. \bseriesno{2}.
\bpublisher{Springer},
\blocation{Dordrecht}
(\byear{2013})
\end{bbook}
\endbibitem

\bibitem{JonasAdler2017}
\begin{botherref}
\oauthor{\bsnm{Adler}, \binits{J.}},
\oauthor{\bsnm{Kohr}, \binits{H.}},
\oauthor{\bsnm{{\"O}ktem}, \binits{O.}}:
Operator Discretization Library (ODL).
\url{https://github.com/odlgroup/odl}
\end{botherref}
\endbibitem

\bibitem{WangBovik2004}
\begin{barticle}
\bauthor{\bsnm{Wang}, \binits{Z.}},
\bauthor{\bsnm{Bovik}, \binits{A.C.}},
\bauthor{\bsnm{Sheikh}, \binits{H.}},
\bauthor{\bsnm{Simoncelli}, \binits{E.}}:
\batitle{Image quality assessment: From error visibility to structural similarity}.
\bjtitle{IEEE Transactions on Image Processing}
\bvolume{13},
\bfpage{600}--\blpage{612}
(\byear{2004}).
\doiurl{10.1109/TIP.2003.819861}
\end{barticle}
\endbibitem

\bibitem{WangBovik2009}
\begin{barticle}
\bauthor{\bsnm{Wang}, \binits{Z.}},
\bauthor{\bsnm{Bovik}, \binits{A.C.}}:
\batitle{Mean squared error: Love it or leave it? {A} new look at signal fidelity measures}.
\bjtitle{IEEE Signal Processing Magazine}
\bvolume{26}(\bissue{1}),
\bfpage{98}--\blpage{117}
(\byear{2009}).
\doiurl{10.1109/MSP.2008.930649}
\end{barticle}
\endbibitem

\bibitem{PamNou2015}
\begin{bchapter}
\bauthor{\bsnm{Pambrun}, \binits{J.-F.}},
\bauthor{\bsnm{Noumeir}, \binits{R.}}:
\bctitle{Limitations of the {SSIM} quality metric in the context of diagnostic imaging}.
(\byear{2015}).
\doiurl{10.1109/ICIP.2015.7351345}
\end{bchapter}
\endbibitem

\bibitem{LinChen2022}
\begin{barticle}
\bauthor{\bsnm{Lin}, \binits{L.}},
\bauthor{\bsnm{Chen}, \binits{H.}},
\bauthor{\bsnm{Kuruoglu}, \binits{E.}},
\bauthor{\bsnm{Zhou}, \binits{W.}}:
\batitle{Robust structural similarity index measure for images with non-gaussian distortions}.
\bjtitle{Pattern Recognition Letters}
\bvolume{163},
\bfpage{10}--\blpage{16}
(\byear{2022}).
\doiurl{10.1016/j.patrec.2022.09.011}
\end{barticle}
\endbibitem

\bibitem{Zit2003}
\begin{barticle}
\bauthor{\bsnm{Zitová}, \binits{B.}},
\bauthor{\bsnm{Flusser}, \binits{J.}}:
\batitle{Image registration methods: A survey}.
\bjtitle{Image and Vision Computing}
\bvolume{21}(\bissue{11}),
\bfpage{977}--\blpage{1000}
(\byear{2003})
\end{barticle}
\endbibitem

\end{thebibliography}
    
\end{document}